\documentclass[preprint2,times,tighten]{aastex6}
\usepackage{amsmath,amstext}
\usepackage[all]{hypcap} 
\newcommand{\rom}[1]{\uppercase\expandafter{\romannumeral #1\relax}}

\usepackage{gensymb}
\usepackage{graphicx}
\usepackage{amsmath}
\usepackage{amssymb}
\usepackage{natbib}
\usepackage[]{units}
\usepackage{placeins}

\newcommand{\halpha}{H$\alpha$}
\newcommand{\hbeta}{H$\beta$}
\newcommand{\ntwo}{$\left[ \text{\ion{N}{2}} \right]$}
\newcommand{\stwo}{$\left[ \text{\ion{S}{2}} \right]$}

\newcommand{\kms}{km\,s$^{-1}$}

\newcommand{\iha}{$I_{\text{H}\alpha}$}
\newcommand{\isii}{$I_{\text{\stwo}}$}
\newcommand{\tanb}{$\tan{\left|b \right|}$}
\newcommand{\expn}[2]{{#1}\mathrm{e}{#2}}

\shorttitle{WIM in Sag-Car Arm}
\shortauthors{Krishnarao et al.}

\begin{document}

\title{Warm Ionized Medium Throughout the Sagittarius-Carina Arm}
\author{Dhanesh Krishnarao\altaffilmark{1}, L. Matthew Haffner\altaffilmark{1,2}, Robert A. Benjamin\altaffilmark{3}, Alex S. Hill\altaffilmark{4}, Kathleen A. Barger\altaffilmark{5}}
\email{krishnarao@astro.wisc.edu}
\altaffiltext{1}{Department of Astronomy, University of Wisconsin-Madison, Madison, WI, USA}
\altaffiltext{2}{Space Science Institute, Boulder, CO, USA}
\altaffiltext{3}{Department of Physics, University of Wisconsin-Whitewater, Whitewater, WI, USA}
\altaffiltext{4}{Department of Astronomy, Haverford College, 370 Lancaster Ave., Haverford, PA 19041}
\altaffiltext{5}{Department of Physics and Astronomy, Texas Christian University, TCU Box 298840, Fort Worth, TX 76129}

\begin{abstract}
Wisconsin H-Alpha Mapper (WHAM) observations of \halpha~and \stwo~$\lambda6716$ emission are used to trace the vertical distribution and physical conditions of the warm ionized medium (WIM) along the Sagittarius-Carina arm. CO emission, tracing cold molecular gas in the plane of the Galaxy, is used as a guide to isolate \halpha~and \stwo~$\lambda6716$ emission along individual spiral arms. Exponential scale heights of electron density squared (or emission measure) are determined using \halpha~emission above (below) the midplane to be $\unit[330 \pm 80]{pc} \left( \unit[550 \pm 230]{pc} \right)$ along the near Sagittarius arm, $\unit[300 \pm 100]{pc} \left( \unit[250 \pm 30]{pc} \right)$ along the near Carina arm, and $\unit[>1000]{pc}$ along the far Carina arm. The emission measure scale height tends to increase as a function of Galactocentric radius along the Sagittarius-Carina arm for $\unit[R_G > 8]{kpc}$. Physical conditions of the ionized gas are analyzed using the \stwo/\halpha~line ratio, which more closely traces \iha~than height above the plane, z, suggesting a stronger relationship with the in-situ electron density. We interpret this result as further evidence for the majority of the observed diffuse emission originating from in-situ ionized gas as opposed to scattered light from classical \ion{H}{2} regions in the plane.

\end{abstract}

\received{October 26, 2016} \accepted{February 27, 2017}
\maketitle

\section{Introduction}

The warm ionized medium (WIM) is a major component of our Galaxy's interstellar medium (ISM) and is an important tracer of energy transport in star-forming galaxies. Primarily composed of ionized hydrogen, the WIM has a characteristic temperature of $8000$ K, one-third of the surface-mass density of neutral hydrogen (\ion{H}{1}), and a vertical scale height of $h_z \approx 1$ kpc \citep{Haffner2009, Savage2009}. The power requirement of the WIM is equivalent to the total kinetic energy input to the ISM from supernovae \citep{Reynolds1991}. After the existence of a low-density ($\unit[10^{-3}]{cm^{-3}}$) WIM was suggested by \citet{Hoyle1963}, observations of radio pulsar dispersions \citep{Taylor1993,Reynolds1989} and faint optical emission lines \citep{Reynolds1998} have been the primary method of studying this component of the ISM. 

Deep \halpha~imaging shows the existence of extraplanar ionized layers in the disk and halo of other star forming galaxies \citep{Dettmar1990, Rand1990}. Extended WIM layers become less prevalent in galaxies with lower star formation rates, demonstrating the complicated disk-halo connection in galaxies. In normal star-forming galaxies, the WIM component of the ISM accounts for $59\% \pm 19\%$ of the observed \halpha~flux \citep{Oey2007} and $\gtrsim 90\%$ of the H$^+$ mass \citep{Haffner2009}. At larger heights above the plane of galaxies, the WIM becomes increasingly dominant (relative to the cold neutral phase) in addition to the hot ($\approx \unit[10^6]{K}$) phase \citep{Reynolds1991}. 

The primary source of ionization in the WIM is believed to be ionizing radiation from O stars, with photoionization models showing their radiation escaping from \ion{H}{2} regions and following extended paths cleared out through feedback processes and turbulence \citep{Reynolds1990, Ciardi2002, Wood2005, Wood2010}. Although originating from the same ionizing source, physical characteristics of the WIM are distinct from classical \ion{H}{2} regions. Discrete \ion{H}{2} regions in the plane of the Galaxy have a much higher dust content, a much smaller scale height ($\approx \unit[50]{pc}$), and lower temperature ($\approx \unit[6000]{K}$) than the diffuse gas of the WIM \citep{Madsen2006, Kreckel2013}. Additionally, the ionization states of the gas vary in the WIM and classical \ion{H}{2} regions, with the WIM containing mostly ions such as O$^+$ and S$^+$, as opposed to O$^{++}$ and S$^{++}$ in \ion{H}{2} regions \citep{Reynolds1995, Reynolds1998}.

\halpha~emission provides the bulk of the information about the mass and distribution of the WIM, with $\gtrsim 80\%$ of the observed faint \halpha~flux originating in the WIM and $\lesssim 20\%$ from scattered light originating in \ion{H}{2} regions \citep{Reynolds1973, Wood1999, Witt2010, Brandt2012, Barnes2014}. The behavior of the classically forbidden, collisionally excited \stwo~$\lambda6716$ and \ntwo~$\lambda6584$ emission lines (referred to as \stwo~and \ntwo~from hereafter) trace variations in the ionization state and temperature of the emitting gas. 

The \stwo~/ \halpha~and \ntwo~/ \halpha~line ratio both increase with decreasing \halpha~intensity in the Milky Way and other galaxies \citep{Haffner1999, Madsen2006, Haffner2009}, suggesting an increase in the gas temperature at lower gas densities. A positive correlation between these line ratios and height above the plane is also observed in many other galaxies \citep{Domgoergen1997, Otte2002, Hoopes2003}. However, because the \halpha~intensity follows an inverse relationship with height, it is difficult to disentangle the physical cause for change in the observed line ratios, which could be due to changes in gas temperature, gas density, or height above the plane.

Recently, the Wisconsin H-Alpha Mapper (WHAM) has provided an all-sky velocity resolved map of \halpha~emission in the Milky Way \citep{Haffner2003,Haffner2010}. Ongoing multi-wavelength observations of \hbeta, \stwo, \ntwo, and other optical emission lines allow for physical conditions of the WIM to be characterized. In this paper, we use WHAM observations to study the spatial and physical properties of the WIM throughout the Sagittarius-Carina arm. Our vantage point within the Galaxy provides an edge-on perspective to study the vertical structure of the WIM in detail. The analysis performed here is motivated by previous work on the Perseus and Scutum-Centaurus arms \citep{Haffner1999, Hill2014}, but incorporates a novel method of kinematically isolating emission along the spiral arm. 

In Section \ref{obs}, we describe our observations and Section \ref{sec_map} presents our spectroscopic maps of the Sagittarius-Carina arm. Section \ref{vert} details our derivation of the scale height of electron density squared (or emission measure, EM) throughout the spiral arm and Section \ref{ratio_stat} begins to analyze the physical conditions of the WIM within the Carina arm using \stwo~emission line data as a tracer of temperature variations. In Section \ref{disc}, we discuss some of our surprising results for the scale height in the far Carina arm and argue that the bulk of the observed emission corresponds to in-situ emission of photoionized gas. Finally, we wrap up with conclusions and a summary in Section \ref{summary}. 

\section{Observations} \label{obs}

WHAM is a dual-etalon Fabry-Perot spectrometer designed to obtain highly sensitive observations of the WIM. WHAM has a $12$ \kms$~$spectral resolution, a $1\degree$ beam, and observes faint optical emission of \halpha, \stwo, \ntwo, \hbeta, and other lines. Our \halpha$~$observations are from the WHAM Sky Survey, which combines observations taken at the Kitt Peak National Observatory \citep[see][]{Haffner2003} and at the Cerro Tololo Inter-American Observatory \citep[see][]{Haffner2010}. Additionally, we use preliminary WHAM \stwo$~$observations from a survey of the southern Galactic plane \citep{Gotisha2013}. \stwo~observations are only available along the Carina arm direction in the fourth quadrant of the Galaxy.

\halpha$~$and \stwo$~$data are observed using 30-second and 60-second exposures through a $1\degree$ beam, producing an average spectrum of emisison over this area. For both \halpha$~$and \stwo$~$data, we have applied a flat field, an atmospheric template, and a constant baseline to reach a $3\sigma$ sensitivity of $\approx \unit[0.1]{R}$ ($\unit[1]{Rayleigh(R)}~=~\unit[\nicefrac{10^6}{4 \pi}]{photons~s^{-1}~cm^{-2}~sr^{-1}}$; 1 R corresponds to an emission measure of $EM = \unit[2.25]{cm^{-6} pc}$ for $T_e = \unit[8000]{K}$). The \halpha$~$ data have had a single Gaussian term subtracted corresponding to the geocoronal \halpha$~$emission line. No corrections for dust extinction have been applied to these data; analyzed sight-lines are restricted to Galactic latitudes $b > 5\degree$. Details on the instrument, its design, observation modes, and data reduction methods are described in \citet{Haffner2003}.


\section{Isolating Spiral Arm Emission} \label{sec_map}
\begin{figure}[htb!]
\label{gal_map}
\epsscale{1.25}
\plotone{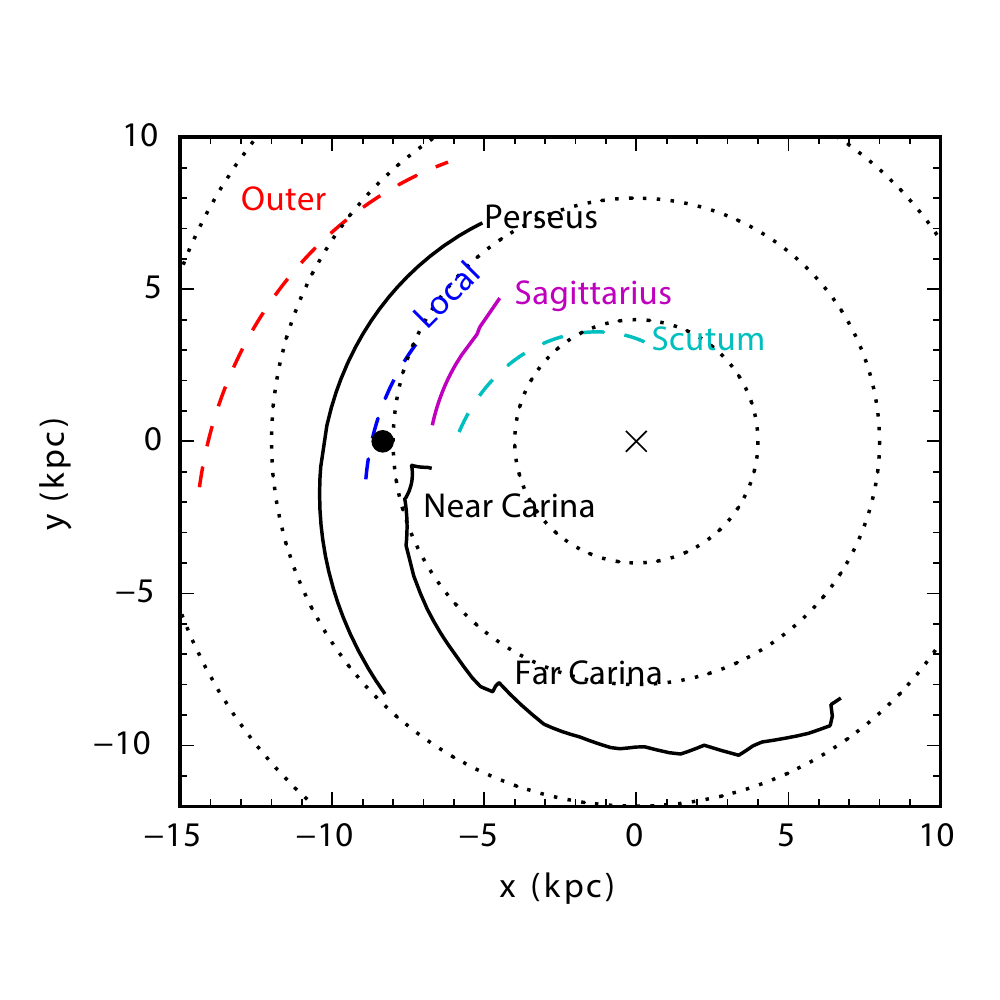}
\caption{Spiral structure model of the Milky Way. Galactic center (black cross) is at ($\unit[0]{kpc}$, $\unit[0]{kpc}$) with circles (black dotted lines) spaced at $4$, $8$, $12$, and $16$ kpc. The Sun (solid black circle) is at ($\unit[-8.34]{kpc}$, $\unit[0]{kpc}$) and Galactic rotation is in the clockwise direction. Dashed lines are logarithmic spiral arm fits for the Outer arm (red), Perseus arm (yellow), Local arm (blue), Sagittarius arm (magenta), and Scutum arm (cyan) from \citet{Reid2014}. Solid black lines are the spiral arm positions used in this work based on CO emission and the longitude-velocity tracks defined by \citet{Reid2016}. This work assumes the Sagittarius and Carina arms are connected and part of the same coherent spiral structure (also referred to as the Sagittarius-Carina arm).}
\end{figure}

\begin{figure*}[htb!]
\label{co}
\epsscale{1.1}
\plotone{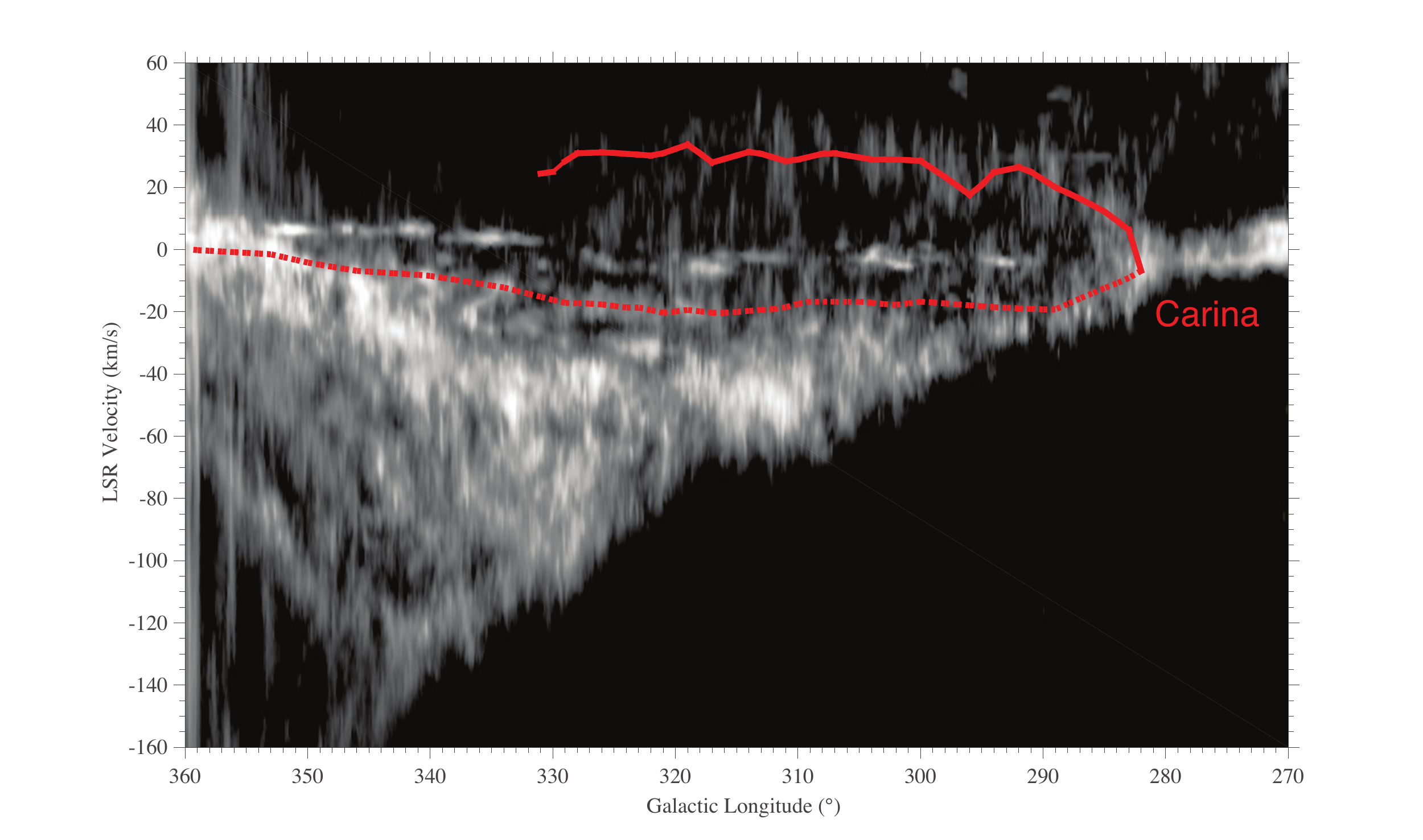}
\caption{ Figure reproduced from Figure 13 of \citet{Reid2016}. CfA CO survey longitude-velocity diagram integrated from $b = +5\degree$ to $-5\degree$ showing a trace of the Carina arm (red). The solid and dotted lines trace far and near portions of the structure, respectively.}
\end{figure*}

Galactic longitude-velocity diagrams of neutral hydrogen and molecular gas emission have long been the traditional method for identifying spiral structure in the Milky Way \citep{Weaver1970, Cohen1985, Dame2007, Reid2014}. However, moving from velocity space into physical distances is not straightforward and requires many assumptions. Figure \ref{gal_map} displays the idealized spiral structure model used in this work as viewed from the north Galactic pole. Molecular gas, as traced by CO emission, is used to define tracks of peak emission in longitude-velocity space (frequently called longitude-velocity, or l-v diagrams) for different Galactic structure features in \citet[][see Appendix]{Reid2016}, as reproduced in Figure \ref{co_sag} and Figure \ref{co}. Their work also provides detailed information on distances and Galactocentric radii along these longitude-velocity tracks where maser parallaxes are used to constrain the distances. 

\begin{figure*}[htb!]
\label{co_sag}
\epsscale{.9}
\plotone{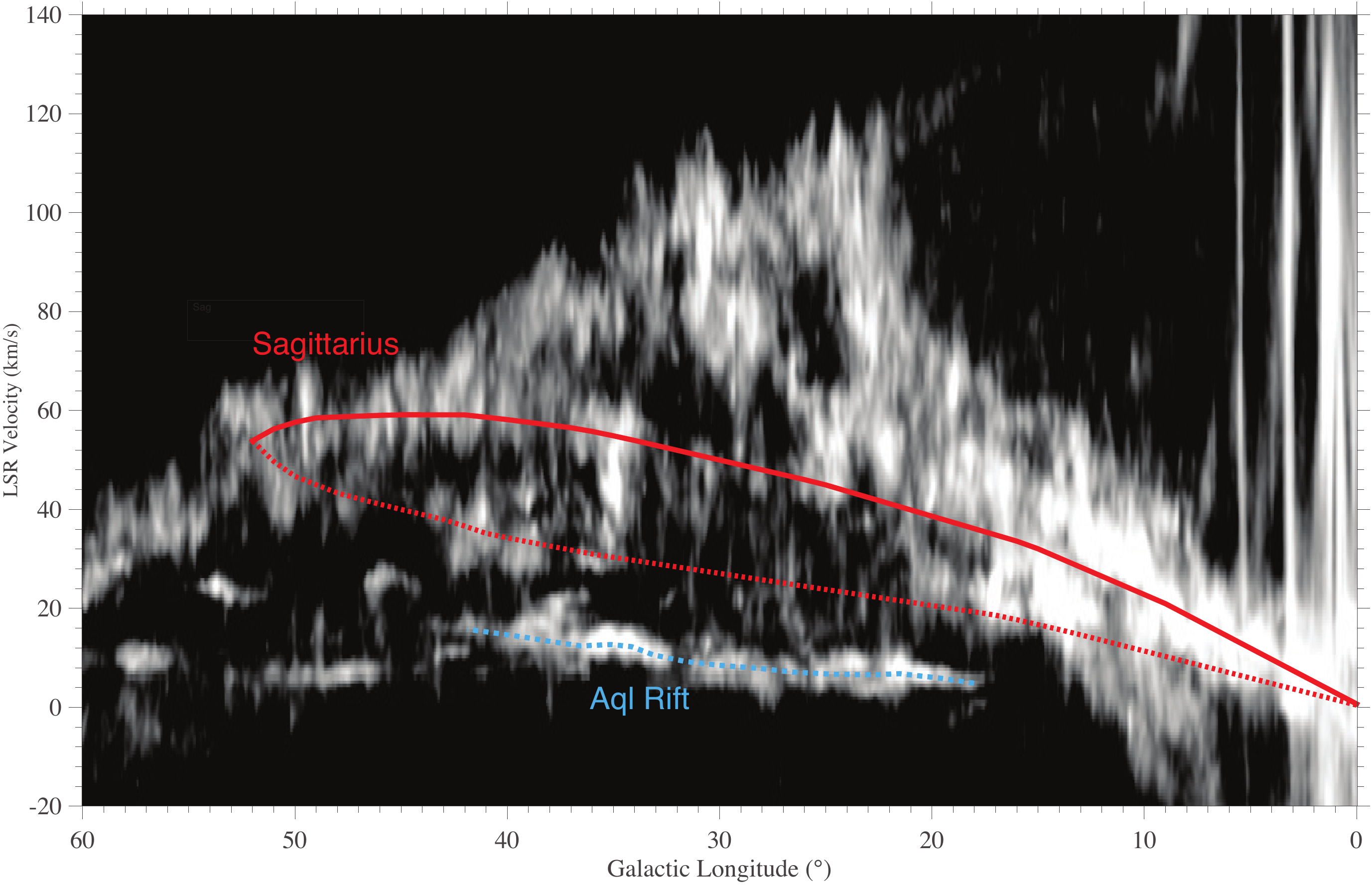}
\caption{Figure reproduced from Figure 7 of \citet{Reid2016}. CfA CO survey longitude-velocity diagram integrated from $b = +1\degree$ to $-1\degree$ showing traces of the Sagittarius arm (red) and the Aquila Rift (blue). The solid and dotted lines trace far and near portions of the structure, respectively.}
\end{figure*}

In this work, these spiral arm tracks are taken as a standard and use kinematic distances assuming $v_{circ} = \unit[220]{km s^{-1}}$ and $R_{\odot} = \unit[8.34]{kpc}$ when a parallax based distance constraint is not given. Since the longitude-velocity tracks are defined solely using observed emission of CO and H\rom{1} gas, purely kinematic distances do not convert into a smooth spiral shape, but rather a jagged pattern as seen along the near and far Carina arm in Figure \ref{gal_map}. For more details on how longitude-velocity diagrams can be used to derive spiral structure in the Milky Way, see \citet[][Section 2 and 3]{Weaver1970}. 

\begin{figure}[htb!]
\label{spectra}
\epsscale{1.15}
\plotone{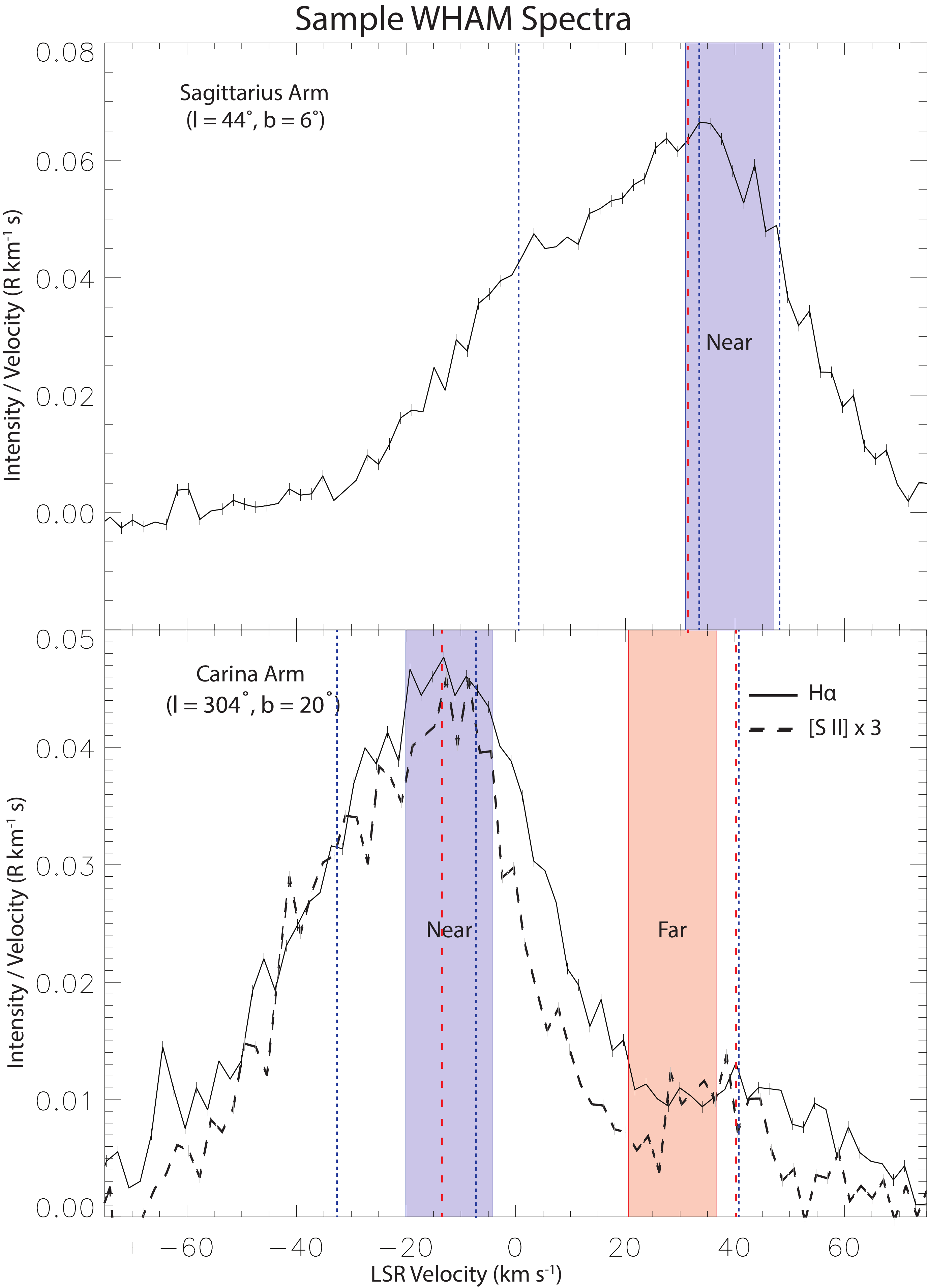}
\caption{Representative spectra as observed with WHAM along a line of sight toward the Sagittarius arm (upper panel) and the Carina arm (lower panel). The solid line shows \halpha~emission and the dashed line shows \stwo~emission multiplied by a factor of three. The shaded regions enclose the integrated region of the spectra used to isolate emission from the near (blue) and far (red) portions of the spiral arm (see Section \ref{sec_map}, Figures \ref{maps} and \ref{mapss} for the the resulting channel maps). Blue dotted lines show the location of the fit Gaussian peaks and red dashed lines show the location of the local maxima (see Figure \ref{bv_peak}). An offset in the peak velocity of \halpha~emission for the far Carina arm from the peak CO emission velocity (red shaded region) is seen (see Section \ref{disc_far}.}
\end{figure}

\begin{figure}[htb!]
\label{bv_peak}
\epsscale{1.2}
\plotone{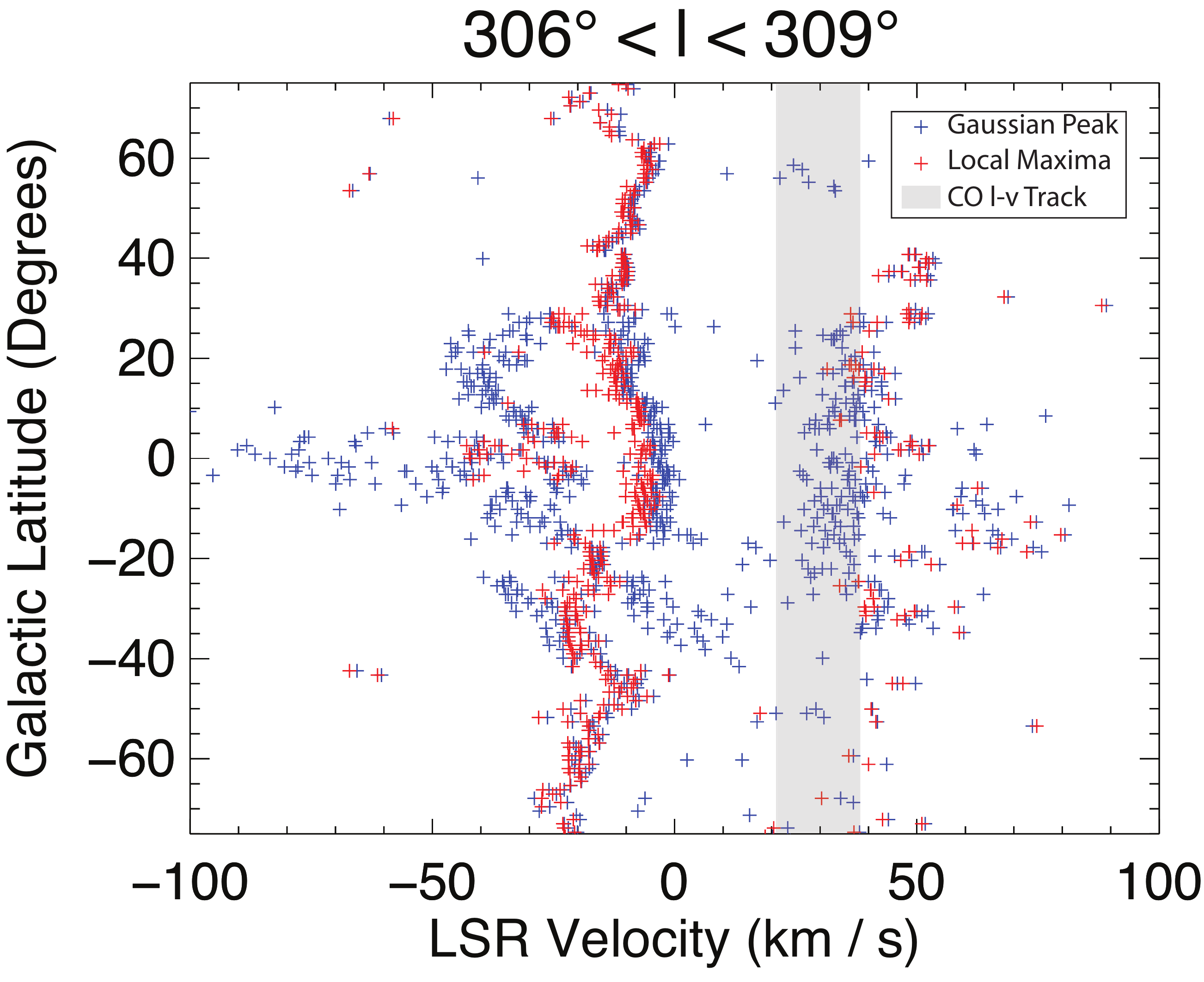}
\caption{Galactic latitude as a function of LSR velocity for WHAM Gaussian peaks. Blue points represent Gaussian peaks manually fit to the data during the atmospheric subtraction process. Red points represent local maxima in the observed spectra. The shaded vertical box encloses the velocity window that data are integrated over for this longitude slice (see Section \ref{sec_map}, Figures \ref{maps} and \ref{mapss} for the the resulting channel maps). Some positive velocity peaks in emission are not covered by this CO l-v track (see Section \ref{disc_far}).}
\end{figure}



The presence of the far Carina arm in WHAM observations was first noticed in the spectra as seen in Figure \ref{spectra}. The presence of faint emission at positive local standard of rest (LSR) velocities led us to investigate the data through individual Gaussian components of the spectra as seen in Figure \ref{bv_peak}. The Gaussian components, shown as blue points, are those fit to the data during the reduction process. Red points are local maxima in the spectra. The collection of points at positive velocities along these longitudes suggest a detection of ionized gas at galactocentric radii of $R_G > R_{\odot} = \unit[8.34]{kpc}$ and closely corresponds with the CO longitude-velocity (l-v) track of the far Carina arm. This relationship led us to use CO emission as a guide to kinematically isolate the \halpha~emission from the spiral arm as a function of longitude. This method separates emission from the near and far portions of the Carina arm along the same line of sight. 

\begin{figure}[htb!]
\label{sag_map}
\epsscale{1.2}
\plotone{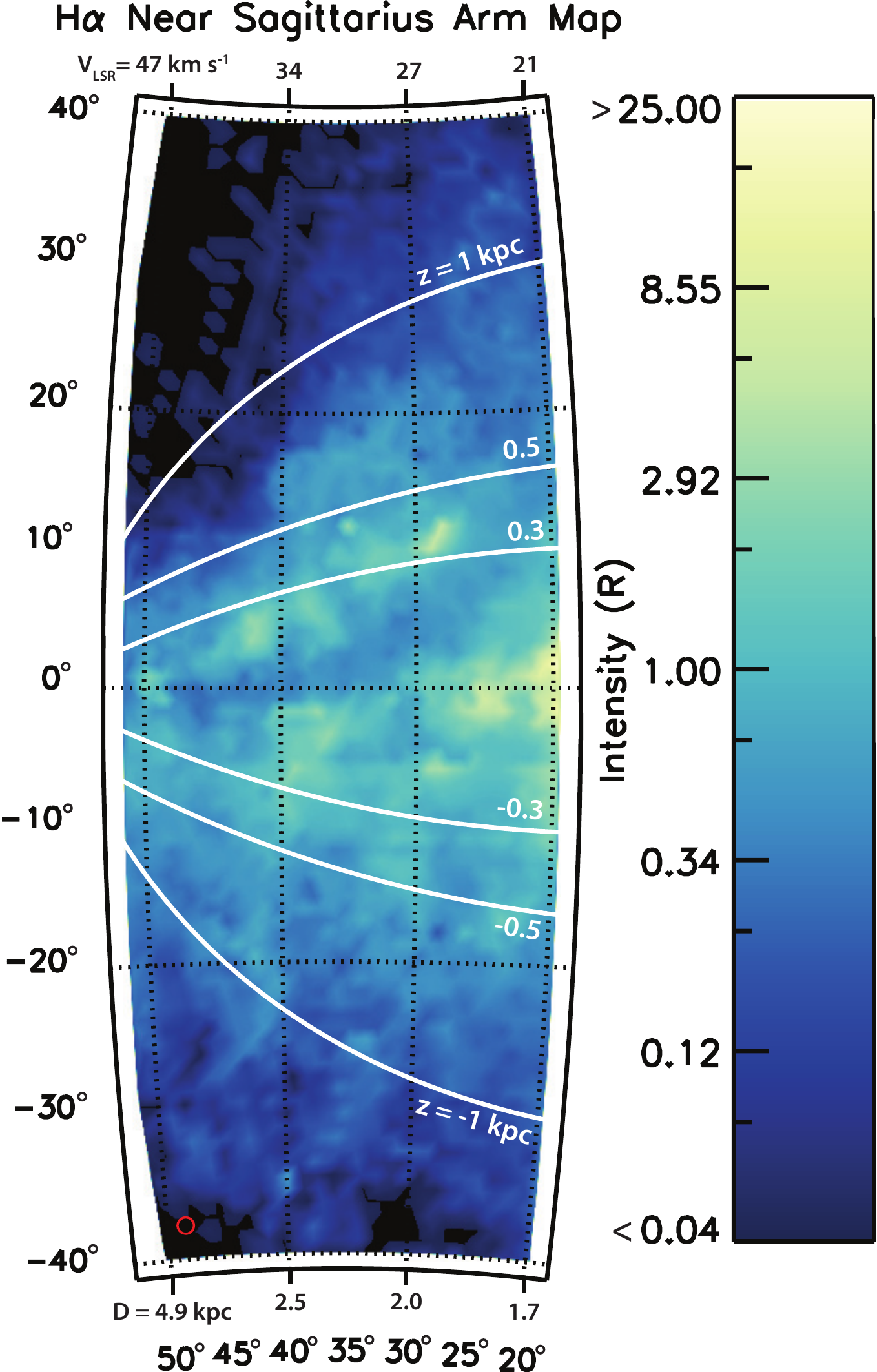}
\caption{Smoothed map of \iha~along the near Sagittarius arm in Galactic coordinates. Data are integrated over a $\unit[16]{km~s^{-1}}$ window centered around the CO l-v track from \citet{Reid2016}, reproduced in Figure \ref{co_sag} and Figure \ref{co}. The solid white lines show different levels of constant height, $z$, above and below the midplane at the assumed distances for the spiral arm structure. Central CO-informed velocities are shown along the upper axis and assumed distances are shown along the lower axis. This direction of the sky shows strong extinction from the Aquila Rift, as seen by the drop in \iha$~$near the midplane. The red circle shows the size of the $1\degree$ WHAM beam.}
\end{figure}
\begin{figure*}[htb!]
\label{maps}
\epsscale{1.1}
\plotone{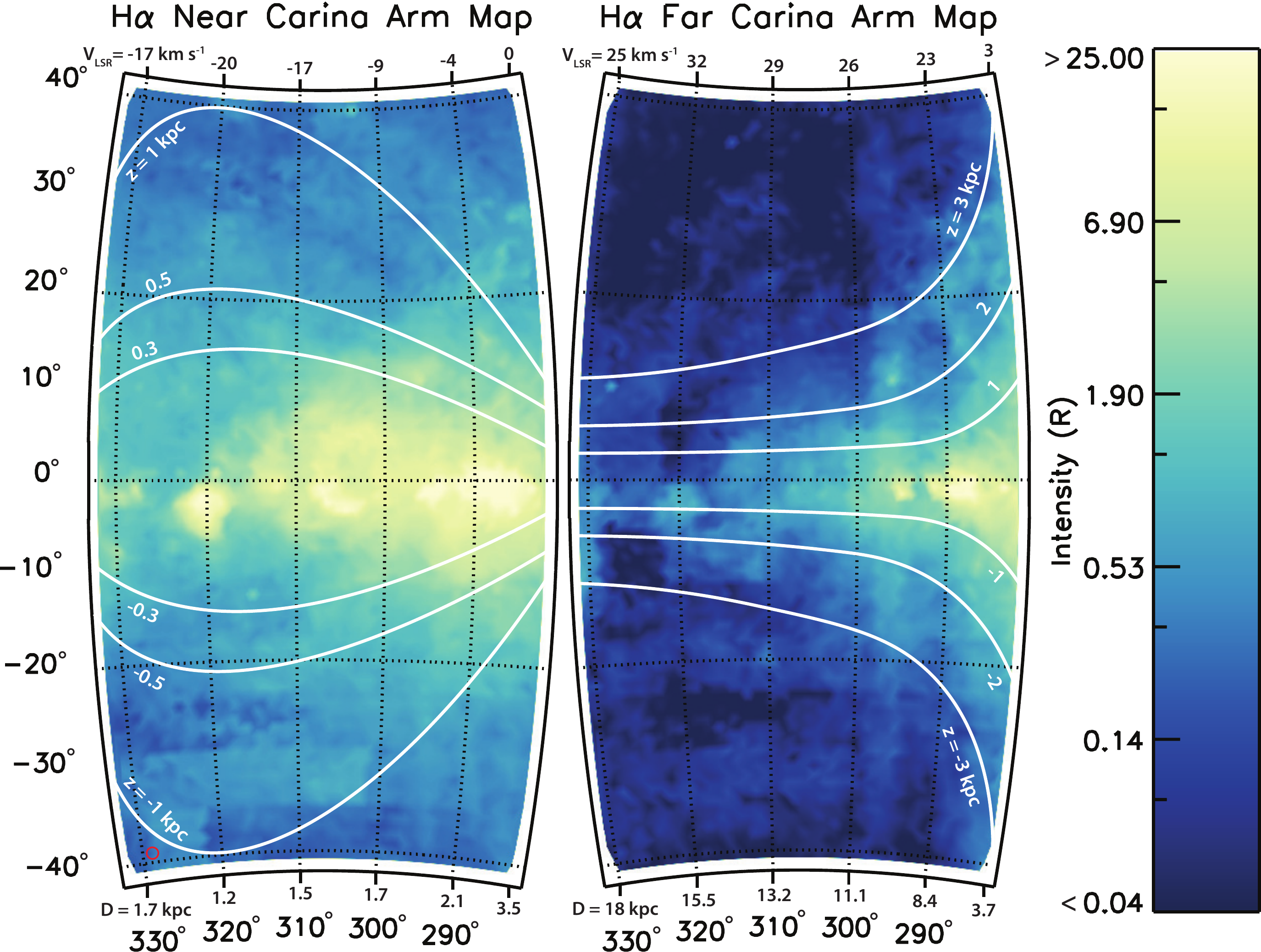}
\caption{Smoothed maps of \iha~throughout the near (left) and far (right) portions of the Carina arm in Galactic coordinates. Data are integrated over a $\unit[16]{km~s^{-1}}$ window centered around the CO l-v track from \citet{Reid2016}, reproduced in Figure \ref{co_sag} and Figure \ref{co}.  The solid white lines show different levels of constant height, $z$, above and below the midplane at the assumed distances for the spiral arm. Central-CO informed velocities are shown along the upper axis and assumed distances are shown along the lower axis. The red circle shows the size of the $1\degree$ WHAM beam.}
\end{figure*}
\begin{figure*}[htb!]
\label{mapss}
\epsscale{1.1}
\plotone{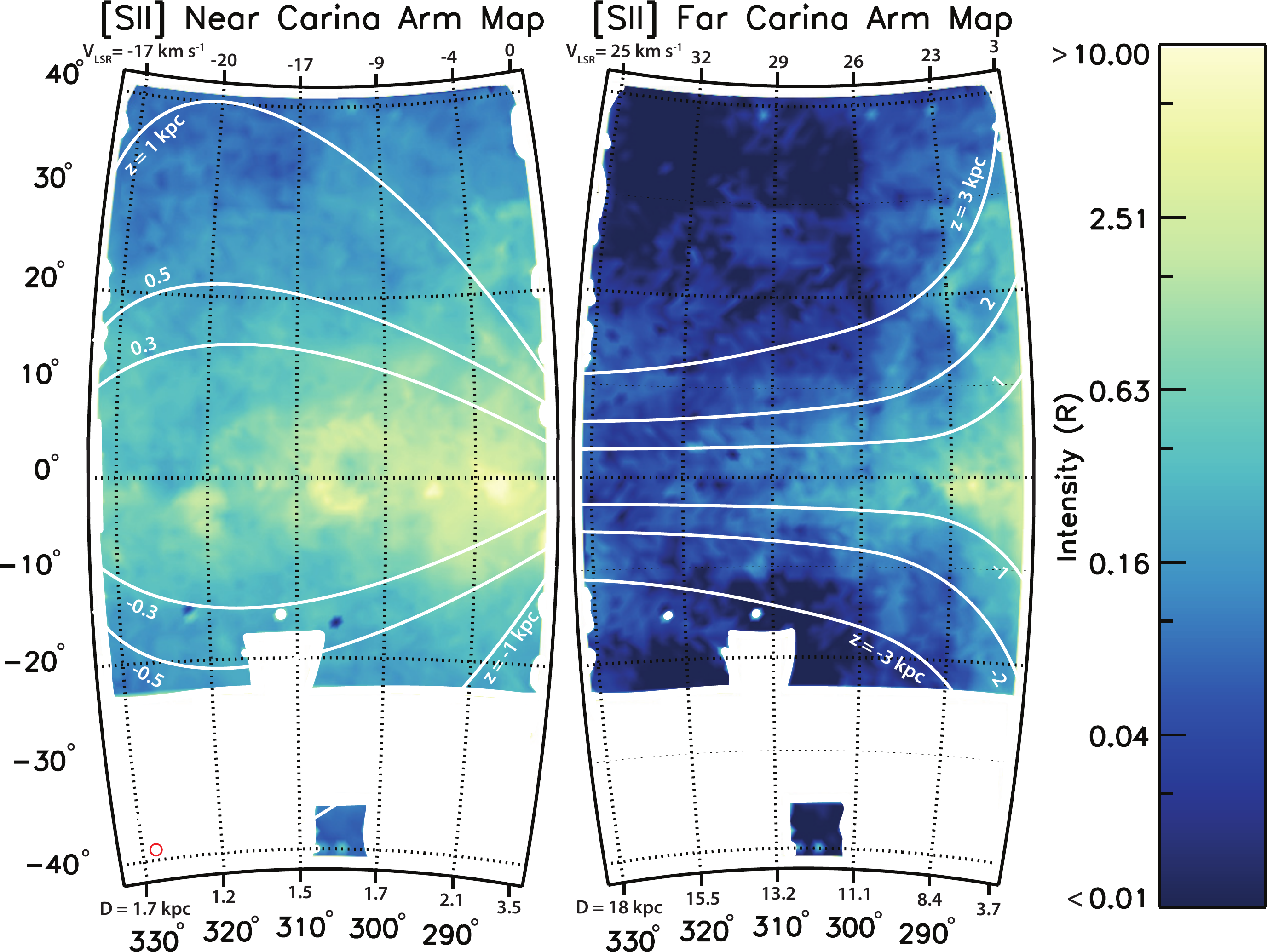}
\caption{Smoothed maps of \isii~throughout the near (left) and far (right) portions of the Carina arm in Galactic coordinates. Data are integrated over a $\unit[16]{km~s^{-1}}$ window centered around the CO l-v track from \citet{Reid2016}, reproduced in Figure \ref{co_sag} and Figure \ref{co}. The solid white lines show different levels of constant height, $z$, above and below the midplane at the assumed distances for the spiral arm. Central CO-informed velocities are shown along the upper axis and assumed distances are shown along the lower axis. The red circle shows the size of the $1\degree$ WHAM beam.}
\end{figure*}
\begin{figure*}[htb!]
\label{ratiomap}
\epsscale{1.15}
\plotone{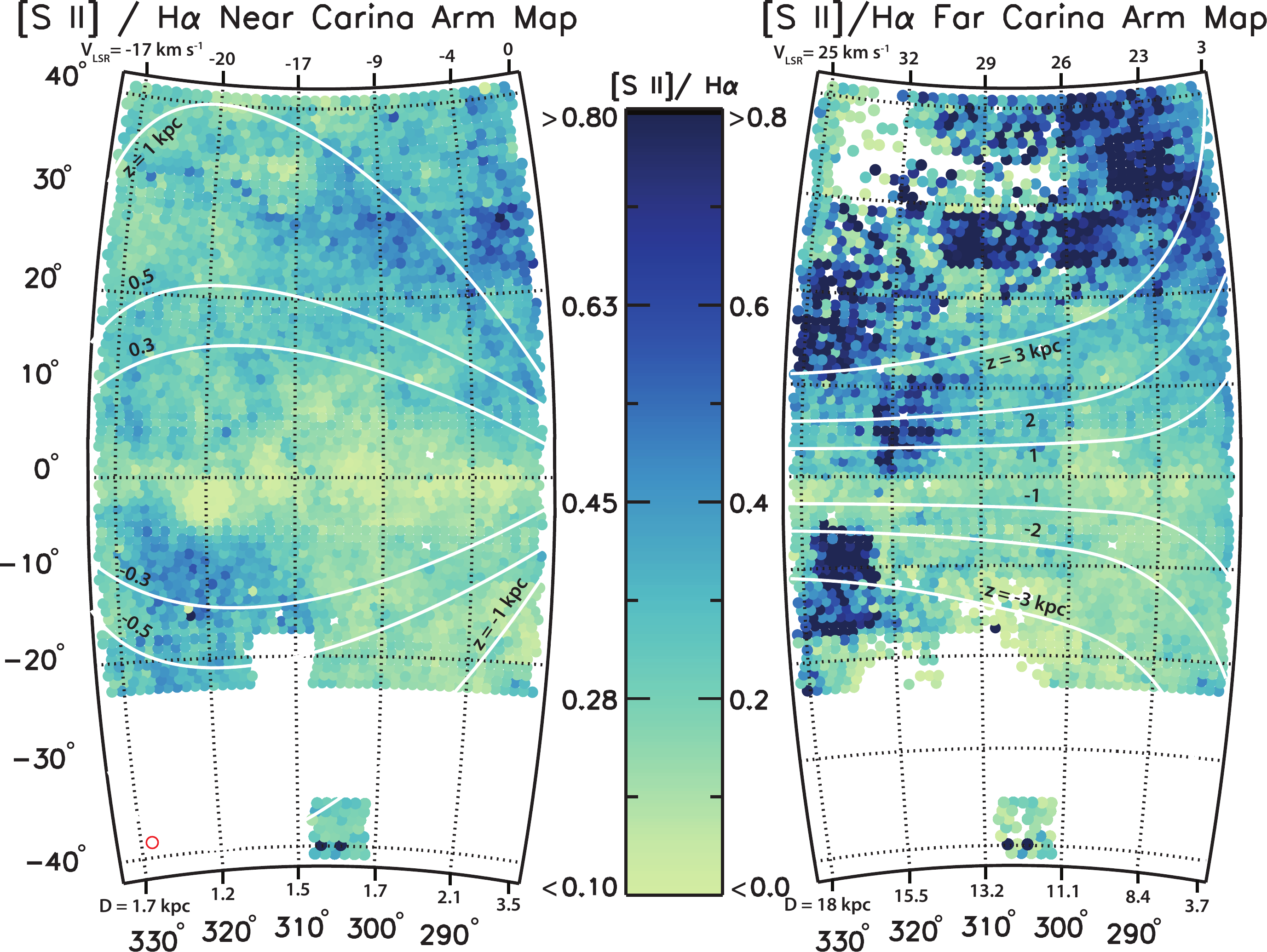}
\caption{Map of the \stwo$~$/ \halpha$~$line ratio in Galactic coordinates throughout the near (left) and far (right) portions of the Carina arm. The solid white lines show different levels of constant height, $z$, above and below the midplane at the assumed distances for the spiral arm. Central CO-informed velocities are shown along the upper axis and assumed distances are shown along the lower axis. The red circle shows the size of the $1\degree$ WHAM beam.} 
\end{figure*}

WHAM data is integrated over a $\unit[16]{km~s^{-1}}$ window centered around the CO l-v tracks for the Sagittarius-Carina arm \citep[see Figure \ref{co_sag},  Figure \ref{co} and][]{Reid2016}. The $\unit[16]{km~s^{-1}}$ width selects peaks of emission rather than encompassing full Galactic emission features, which typically have a width around $\unit[20 - 30]{km~s^{-1}}$. The narrow width better separates the arm emission from local sources ($v_{LSR} \approx \unit[0]{km~s^{-1}}$). Figure \ref{sag_map} shows the velocity-channel map of \iha~for the near Sagittarius arm ($20\degree < l < 52\degree$). Velocity-channel maps of \iha, \isii, and the line ratio \stwo$~$/ \halpha$~$for the near and far portions of the Carina arm are in Figures \ref{maps}, \ref{mapss}, and \ref{ratiomap} ($282\degree < l < 332\degree$).

The far Carina arm shows a strong perspective effect in \halpha~and \stwo~emission, as the spiral arm increases in distance with increasing Galactic longitude (see Figure \ref{maps} and Figure \ref{mapss}). Our confidence in observing this distant spiral arm is explained in detail in Section \ref{disc_far}. The near Sagittarius arm also shows a perspective effect as the distance to the arm segment also increases with Galactic longitude (see Figure \ref{sag_map}). Significant extinction from the Aquila Rift is seen across the near Sagittarius arm near the midplane. 

The map of \isii$~$for the Carina arm is incomplete, and observations are in progress for $b \lesssim -20\degree$ and for other portions of the sky. The points in the \stwo~/ \halpha~line ratio maps show the size of the $1\degree$ WHAM beam. The line ratio \stwo$~$/ \halpha$~$along the far Carina map seems to follow the same perspective effect. The \stwo$~$/ \halpha$~$line ratio generally increases with height above the midplane (see Section \ref{ratio_stat}). 

\section{Vertical Extent of the WIM} \label{vert}

\begin{figure}[htb!]
\label{fits_2}
\epsscale{1.15}
\plotone{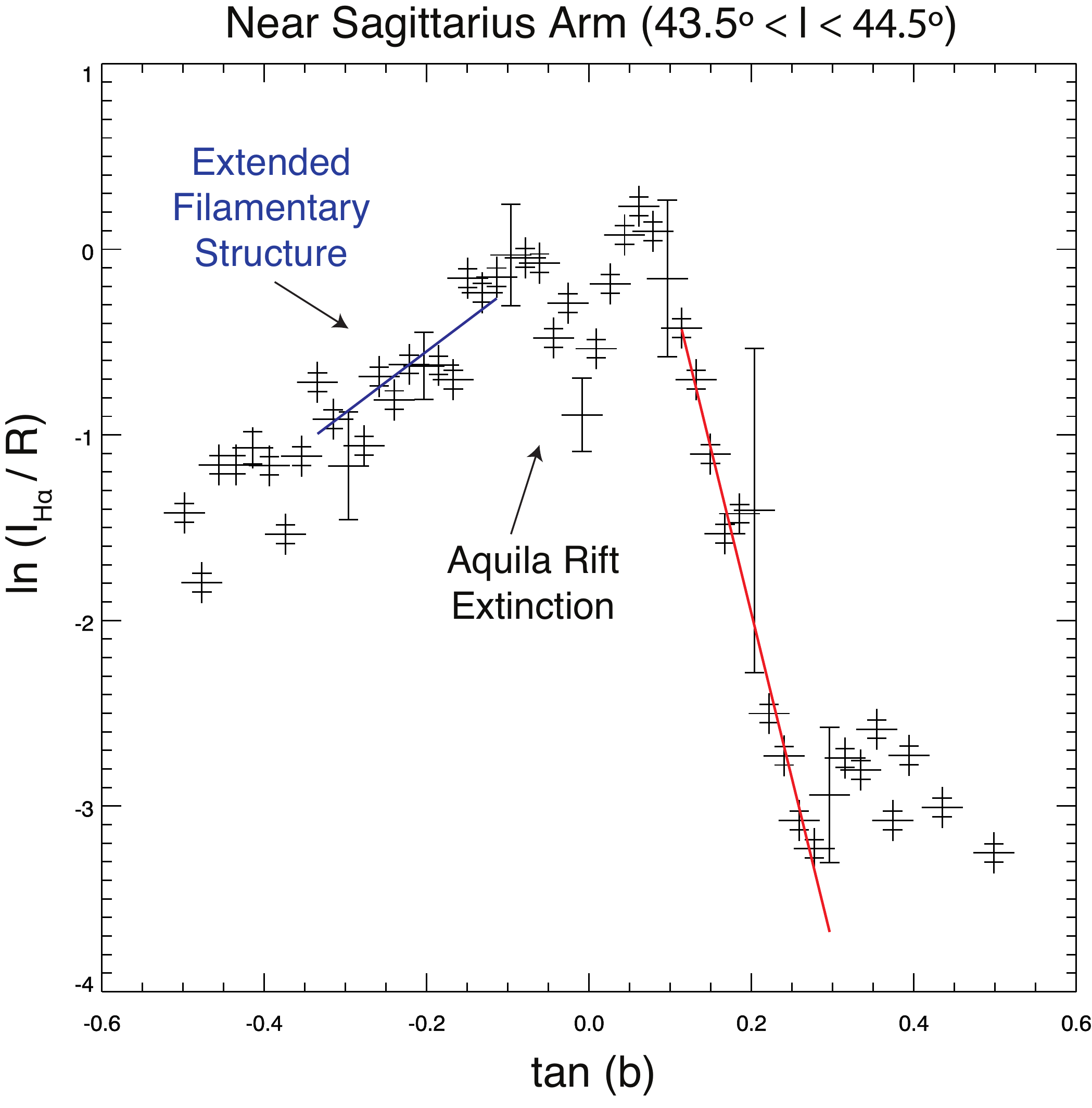}
\caption{\halpha~intensity as a function of $\tan{\left(b\right)}$ between $43.5\degree < l < 44.5\degree$ for the near Sagittarius arm. The red and blue lines show fits of Equation \ref{hfit} above and below the plane, respectively. Note the large extinction feature near $b = 0\degree$ caused by the Aquila Rift. The significantly different profiles above and below the plane show how the filamentary structure is extending far below the midplane.}
\end{figure}

\begin{figure}[htb!]
\label{fits}
\epsscale{1.15}
\plotone{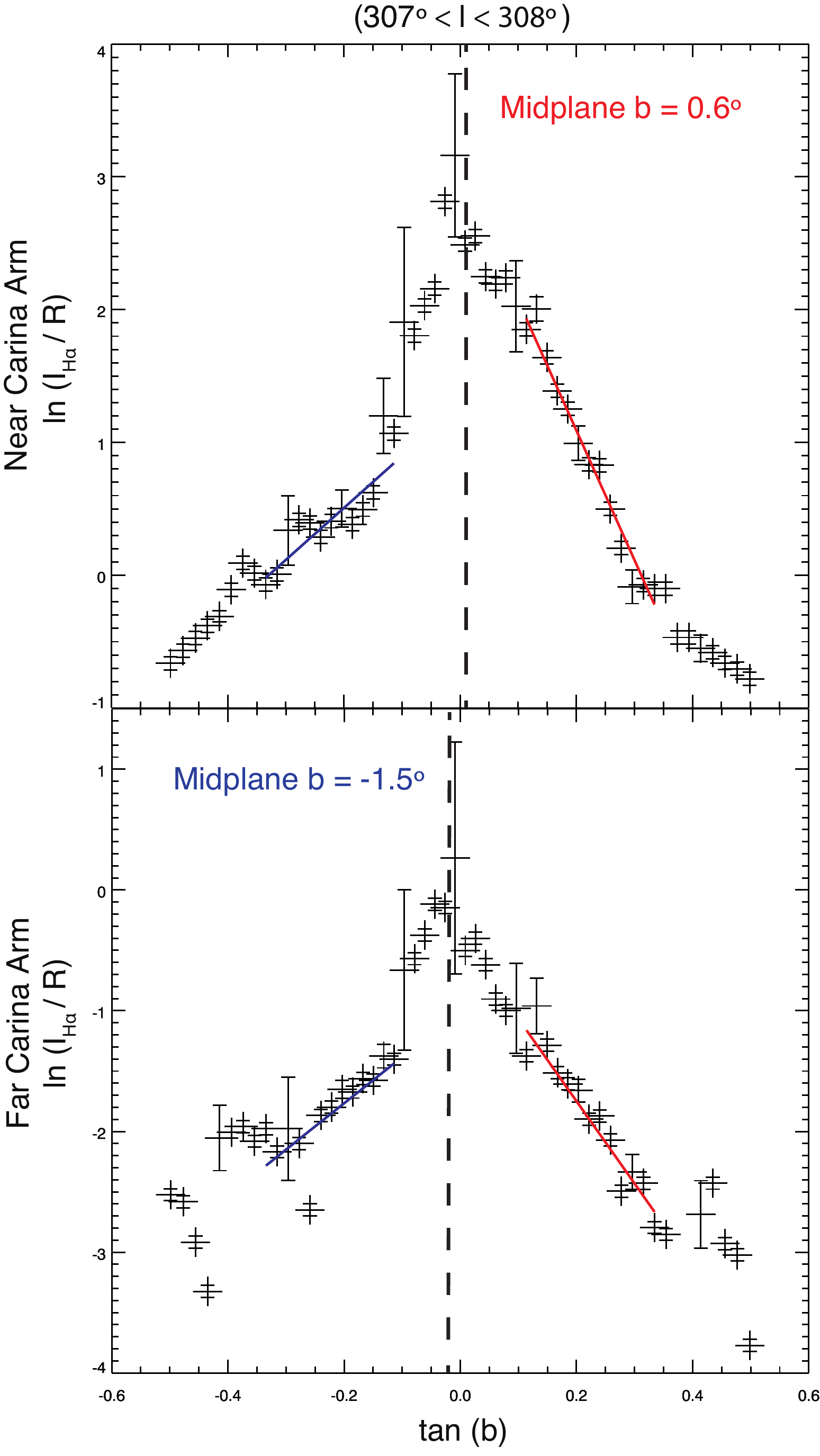}
\caption{\halpha~intensity as a function of $\tan{\left(b\right)}$ between $307\degree < l < 308\degree$ for the near (top) and far (bottom) Carina arm. The red and blue lines show fits of Equation \ref{hfit} above and below the plane, respectively. The dashed line marks the location of the midplane, as determined by fitting a single Gaussian to the data. Note the lack of significant extinction near the midplane.}
\end{figure}

We assume \iha, tracing density of electrons squared ($n_e^2$, see Section \ref{v_dis}), follows an exponential drop with height above the midplane, 

\begin{equation}
n_e^2 \left(z\right) = \left(n_e^2\right)_0 \exp{\left(- \frac{\left|z\right|}{H_{n_e^2}}\right)}
 \label{expha}
\end{equation}

\noindent where $H_{n_e^2}$ is the scale height of the electron density squared (or EM scale height) and $z$ is the height above the midplane ($H_{n_e} = 2 H_{n_e^2}$ for a constant temperature and filling fraction; see Section \ref{v_dis}). Figure \ref{fits_2} and Figure \ref{fits} show a sample of observed \iha$~$along a fixed Galactic longitude as a function of Galactic latitude. Following \citet{Haffner1999}, we fit

\begin{equation}
\ln{I_\text{\halpha}} = \ln{I_0} - \frac{D}{H_{n_e^2}} \tan{\left|b\right|}
 \label{hfit}
\end{equation}
to the data, where $I_0$ is the midplane intensity (at $b = 0\degree$),  $D$ is the distance to the arm in the midplane, and  $\tan{\left(b\right)}$ is the tangent of the Galactic latitude ($\left|z\right| = D \tan{\left|b\right|}$). The slope of the data, shown in Figure \ref{fits_2} and Figure \ref{fits}, is a direct measure of $D / H_{n_e^2}$. Fits are constrained to Galactic latitudes $\left| b \right| \gtrsim 5\degree$ to leave out \ion{H}{2} regions in the plane. The range in Galactic latitudes are allowed to have slight variations while fitting to account for a shifting midplane (relative to $b = 0\degree$) and to mask local sources of emission. 

Distances and Galactocentric radii estimates to the Sagittarius arm are from \citet{Reid2016}, where constraints are made using parallax measurements for masers. The Carina arm does not have parallax-based distance constraints, so kinematic distances are used assuming $v_{circ} = \unit[220]{km s^{-1}}$ and $R_{\odot} = \unit[8.34]{kpc}$, slightly adjusted to the Bayesian distance estimates from \citet{Reid2016}.  Distance uncertainties for the Carina Arm are inherently large, and these systematic errors are not considered in the rest of our analysis. 

The following sections show derived EM scale heights for the near Sagittarius arm, and the near and far portions of the Carina arm (see Figures \ref{sag}, \ref{near_car}, and \ref{far_car} and Table \ref{sum}). Each data point corresponds to the slope of a $1\degree$ vertical slice of the \iha$~$map along positive (red) and negative (blue) Galactic latitude. Occasional gaps in data points are the result of local contamination or data approaching background noise levels. All error bars are statistical errors on the measured slope and assume no uncertainty in distance. 

\subsection{Near Sagittarius arm} \label{nearsag}

\begin{figure*}[htb!]
\label{sag}
\epsscale{1.15}
\plotone{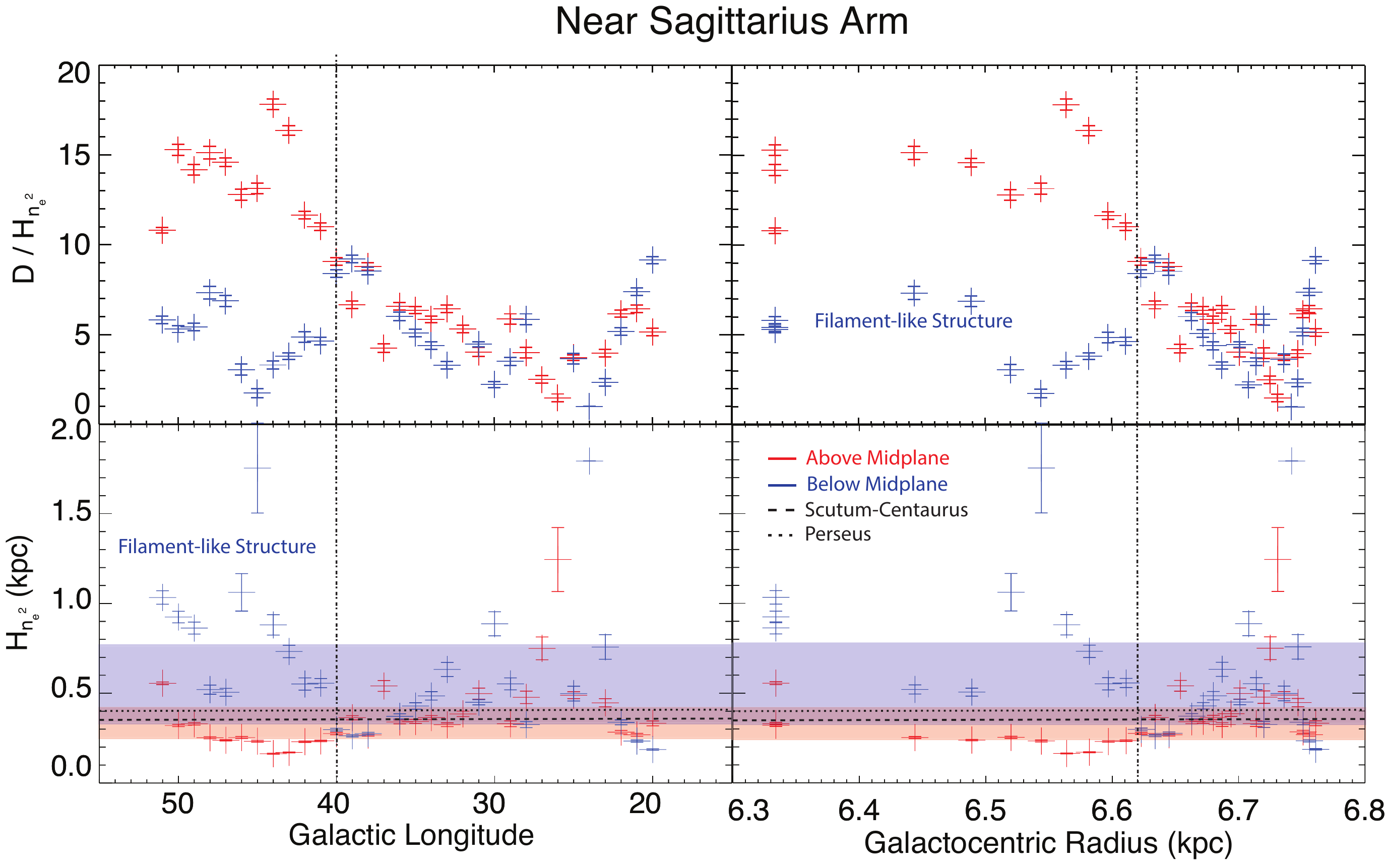}
\caption{Plot showing the measured ratio,$\nicefrac{D}{H_{n_e^2}}$, of distance (D) to the EM scale height ($H_{n_e^2}$), and estimated $H_{n_e^2}$ along the near Sagittarius arm as a function of Galactic longitude and Galactocentric Radius. Red and blue crosses indicate positive and negative Galactic latitudes, respectively. Uncertainties for $\nicefrac{D}{H_{n_e^2}}$ are all smaller than the plotting symbols. All errors here assume zero uncertainty in $D$. The shaded regions show the the median ($\pm$ median absolute deviation from the median) of $H_{n_e^2}$ above (red) and below (blue) the midplane. The dashed and dotted lines represent $H_{n_e^2}$ for the Scutum-Centaurus (extinction-corrected) and Perseus arms, respectively \citep{Hill2014,Haffner1999}. For $l > 40\degree$ ($R_G \lesssim \unit[6.62]{kpc}$), we see evidence for an extended filamentary structure below the plane, resulting in inconsistent measurements above and below the plane. }
\end{figure*}

Results for the near Sagittarius arm are in Figure \ref{sag}, with both $\nicefrac{D}{H_{n_e^2}}$ and $H_{n_e^2}$ as a function of Galactic longitude and Galactocentric radius. Our derived EM scale height generally agrees with the scale heights for the Scutum-Centaurus and Perseus arms, shown as dashed and dotted lines, respectively, in Figure \ref{sag} \citep{Hill2014, Haffner1999}. 

If a constant scale height is assumed along this section, the trend for $\nicefrac{D}{H_{n_e^2}}$ to increase at higher Galactic longitude agrees with known distances to the arm. A drop in the value of $\nicefrac{D}{H_{n_e^2}}$ at larger Galactocentric radii is seen, but the derived EM scale heights do not show this trend. The relationship between $\nicefrac{D}{H_{n_e^2}}$ and Galactocentric radius is likely the result of differences in $D$ to the spiral arm segment. 

Filament-like structures extending towards negative Galactic latitudes from the midplane are seen around $40\degree < l < 46\degree$ and $28\degree < l < 35\degree$. Figure \ref{sag} shows how the negative Galactic latitude slope measurements (in blue) are generally flattened out for $l > 40\degree$ when compared with positive Galactic latitudes (in red). Many of the measured scale height outliers (in blue) correspond to these Galactic longitudes, where this filament-like structure strongly extends the height of the ionized gas below the plane. The filamentary features can also be seen in the full-sky \halpha$~$map from \citet{Finkbeiner2003}, which combined preliminary WHAM velocity-channel maps with higher spatial resolution \halpha$~$imaging from the Virginia Tech Spectral Line Survey \citep[VTSS;][]{Dennison1998} and the Southern H-Alpha Sky Survey Atlas \citep[SHASSA;][]{Gaustad2001}. Further analysis of these filament features is beyond the scope of this paper. 

\subsection{Near Carina Arm}

\begin{figure*}
\label{near_car}
\epsscale{1.15}
\plotone{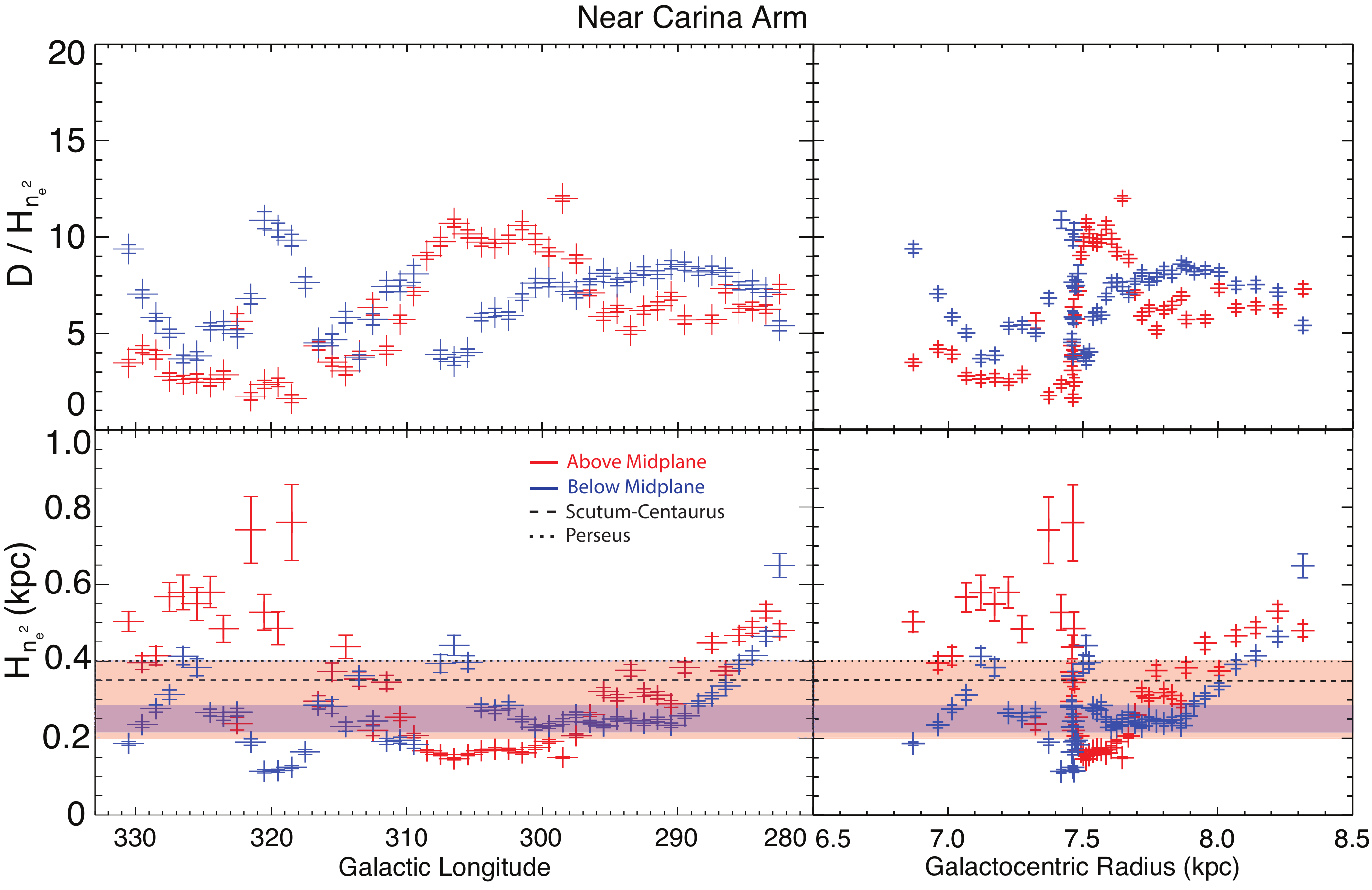}
\caption{Plot showing the measured ratio,$\nicefrac{D}{H_{n_e^2}}$, of distance (D) to the EM scale height ($H_{n_e^2}$), and estimated $H_{n_e^2}$ along the near Carina arm as a function of Galactic longitude and Galactocentric Radius. Red and blue crosses indicate positive and negative Galactic latitudes, respectively. Uncertainties for $\nicefrac{D}{H_{n_e^2}}$ are all smaller than the plotting symbols. All errors here assume zero uncertainty in $D$. Uncertainties in $D$ for the near Carina arm are large and estimates of both $D$ and Galactocentric radii are used. The shaded regions show the the median ($\pm$ median absolute deviation from the median) of $H_{n_e^2}$ above (red) and below (blue) the midplane. The dashed and dotted lines represent $H_{n_e^2}$ for the Scutum-Centaurus and Perseus arms, respectively \citep{Hill2014,Haffner1999}. The gap in points from $315$\degree to $318$\degree is due to an inability to fit a linear regression to the data.}
\end{figure*}

Figure \ref{near_car} shows results for the near Carina arm. The top panel of Figure \ref{near_car} shows $\nicefrac{D}{H_{n_e^2}}$ as directly measured from WHAM, and does not make any assumptions on distance. $\nicefrac{D}{H_{n_e^2}}$ has a minimum value near $l = 320\degree$, corresponding with the expected Galactic longitude of minimum distance to the near Carina arm. Estimated distances are generally low ($\approx \unit[1-2]{kpc} $) and the l-v track is not well separated from local emission at $v_{LSR} = 0$ \kms. Local emission sources and \ion{H}{2} regions dominate the observed emission up to significant Galactic latitudes ($b \approx 10$) and cause some asymmetric measurements about the midplane. However, there is still good agreement, but with larger scatter, with the EM scale height along the Scutum-Centaurus and Perseus arms \citep{Hill2014, Haffner1999}.

\subsection{Far Carina Arm}

\begin{figure*}
\label{far_car}
\epsscale{1.15}
\plotone{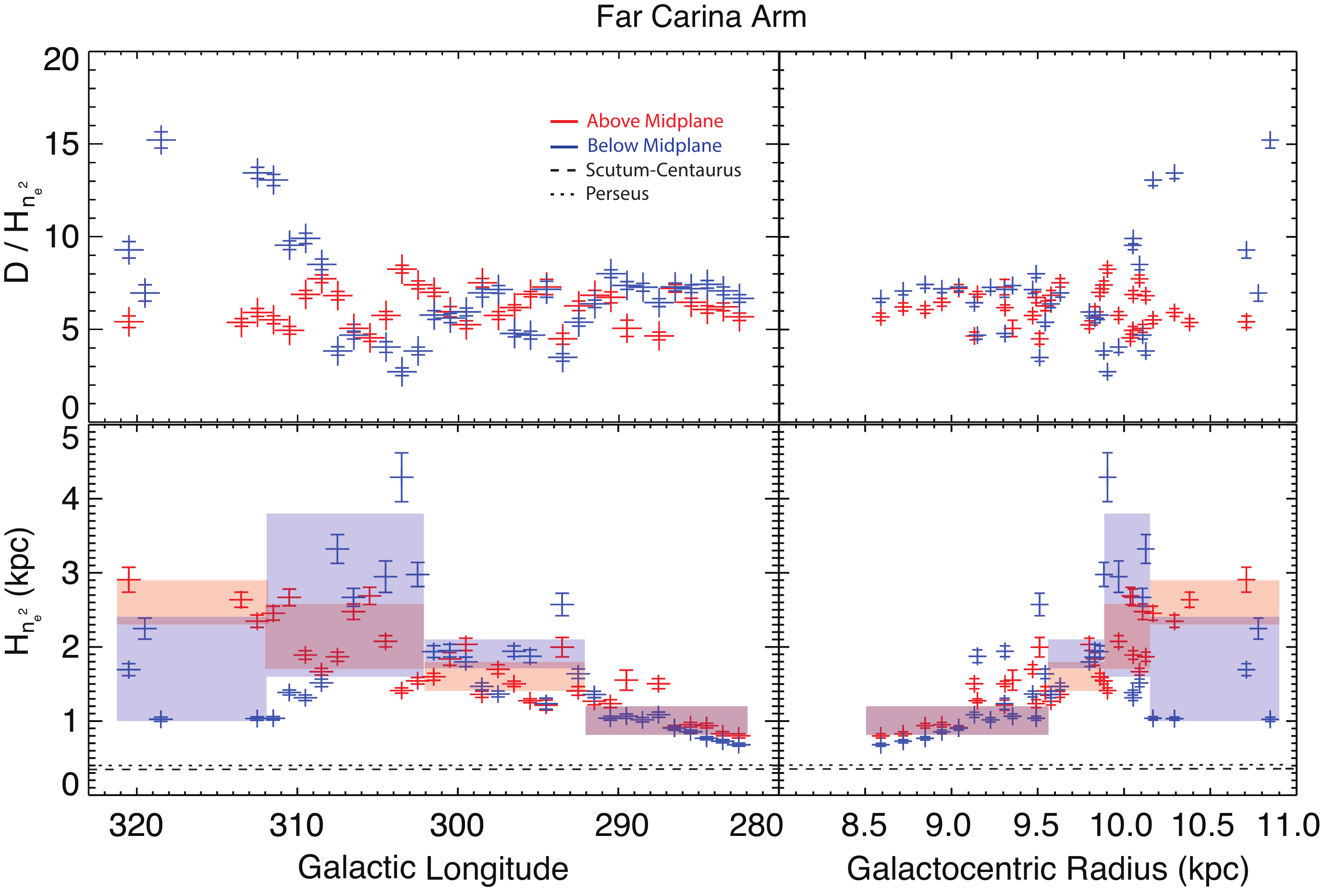}
\caption{Plot showing the measured ratio,$\nicefrac{D}{H_{n_e^2}}$, of distance (D) to the EM scale height ($H_{n_e^2}$), and estimated $H_{n_e^2}$ along the far Carina arm as a function of Galactic longitude and Galactocentric Radius. Red and blue crosses indicate positive and negative Galactic latitudes, respectively. Uncertainties for $\nicefrac{D}{H_{n_e^2}}$ are all smaller than the plotting symbols. All errors here assume zero uncertainty in $D$. Uncertainties in $D$ for the far Carina arm are large and estimates of both $D$ and Galactocentric radii are used. The shaded regions show the the median ($\pm$ median absolute deviation from the median) of $H_{n_e^2}$ above (red) and below (blue) the midplane. The dashed and dotted lines represent $H_{n_e^2}$ for the Scutum-Centaurus and Perseus arms, respectively \citep{Hill2014,Haffner1999}. The gap in points from $315$\degree to $318$\degree is due to an inability to fit a linear regression to the data.}
\end{figure*}

Figure \ref{far_car} shows results for the far Carina arm. $\nicefrac{D}{H_{n_e^2}}$ is roughly constant around $282\degree < l < 310\degree$, and generally larger for $l > 310\degree$, especially at negative Galactic latitudes (blue points). The far Carina arm is increasing in distance as a function of Galactic longitude, resulting in an EM scale height that increases with increasing Galactic longitude and Galactocentric radius. All derived EM scale heights along this far portion of the Carina arm are significantly larger than that of the Scutum-Centaurus and Perseus arms \citep{Hill2014, Haffner1999}. EM scale heights reach up to $H_{n_e^2} \approx \unit[2.5]{kpc}$ for $l > 302\degree$. Near tangency, scale heights are lower, but still much larger than other spiral arm sections, with $H_{n_e^2} \approx \unit[0.5 - 1]{kpc}$. This surprising result is further discussed in Section \ref{disc}.

\subsection{Other Spiral Arms}

Table \ref{sum} and Figure \ref{summary} shows a summary of $\nicefrac{D}{H_{n_e^2}}$ and $H_{n_e^2}$ for the Sagittarius-Carina arm, along with the Scutum-Centaurus, and Perseus arms from \citet{Hill2014} and \citet{Haffner1999}. For the near Sagittarius arm, we report a median value for positive and negative Galactic latitudes for $20\degree < l < 52\degree$. Positive Galactic latitude measurements are likely more representative of the WIM, as the negative side shows filament-like structures at $l > 40\degree$. Median values are shown for the near Carina arm at $286\degree < l < 332\degree$, excluding the tangency region from $282\degree < l < 286\degree$. For the far Carina arm, four median values are shown for $H_{n_e^2}$ across $282\degree<l<292\degree$, $292\degree<l<302\degree$, $302\degree<l<312\degree$, and $312\degree<l<322\degree$. Errors are the median absolute deviation from the median. The scale height for the Perseus arm is modified from measurements made in \citet{Haffner1999} to account for more recent distant constraints (change in distance from $D = \unit[2.5]{kpc}$ to $D = \unit[1.95 \pm 0.04]{kpc}$) \citep[see][]{Hill2014, Xu2006, Reid2014}. A direct measure of $H_{n_e}$ in the local Galaxy as derived from pulsar dispersion measures \citep{Gaensler2008, Savage2009} is also displayed. Assuming a constant filling fraction, $f\left(z\right)$, this value can be converted to a $H_{n_e^2}$, or EM scale height for better comparison (see Section \ref{v_dis} for details on the filling fraction). This local measure of the ionized gas scale height shows good correspondance with the tendency for the ionized gas scale height to increase as a function of Galactocentric Radius along the Sagittarius-Carina arm (see Figure \ref{summary}). 

\begin{figure*}
\label{summary}
\epsscale{1.2}
\plotone{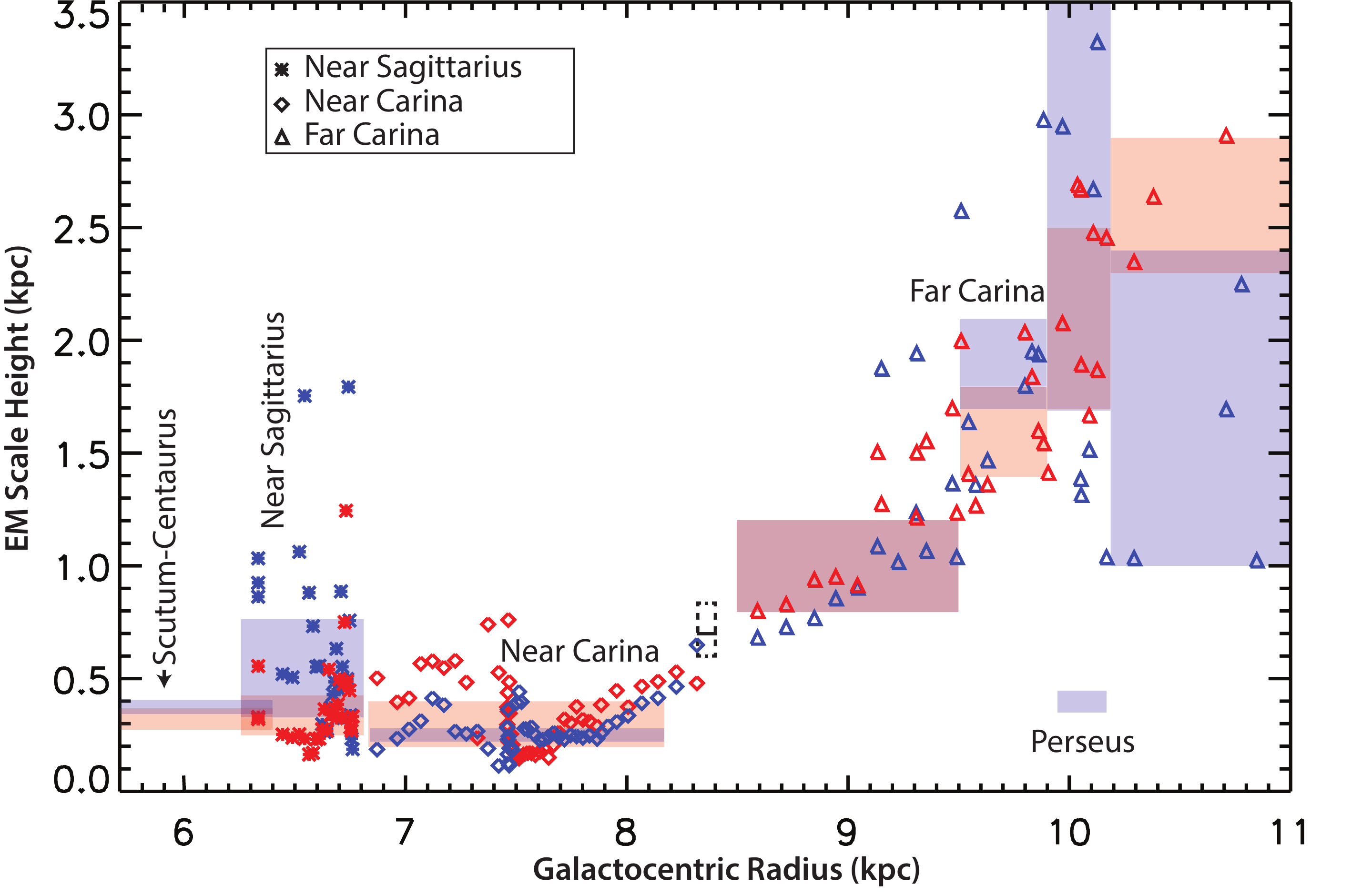}
\caption{Summary of results across the Sagittarius-Carina arm, Scutum-Centaurus arm, and Perseus arm (see Table \ref{sum}). Derived EM scale heights ($H_{n_e^2}$) are plotted for the near Sagittarius arm (asterisk), near Carina arm (diamond), and far Carina arm (triangle) both above (red) and below (blue) the plane. The shaded regions around the data points show the the median ($\pm$ median absolute deviation from the median) of $H_{n_e^2}$. The shaded regions denoted for the Scutum-Centaurus and Perseus arms are reported values from \citet{Hill2014,Haffner1999}. The dashed black box near $R_G = \unit[8.34]{kpc}$ is a local scale height derived from pulsar dispersion measures by \citet{Gaensler2008, Savage2009} converted from an electron scale height ($H_{n_e}$) to an EM scale height ($H_{n_e^2}$) by assuming a constant filling fraction $f\left(z\right)$. }
\end{figure*}

\begin{table}[htb!]
\begin{center}
Near Sagittarius Arm \\
\begin{tabular}{l c c c r }
\hline
$b$			&	$l_{min}$	&	$l_{max}$	&	$\nicefrac{D}{H_{n_e^2}}$	& 	$H_{n_e^2}$ 	\\
($\pm$)		&	(degrees)	&	(degrees)	&							& 	 (pc)			\\
\hline
$+$			& $20\degree$	& $52\degree$	&	$6.6 \pm 2.6$			&	$330 \pm 80$	\\
$-$			& $20\degree$	& $52\degree$	&	$5.1 \pm 1.8$			&	$550 \pm 230$	\\
\hline
\\
\end{tabular}\\
Near Carina Arm \\
\begin{tabular}{l c c c r }
\hline
$b$			&	$l_{min}$	&	$l_{max}$	&	$\nicefrac{D}{H_{n_e^2}}$	& 	$H_{n_e^2}$ 	\\
($\pm$)		&	(degrees)	&	(degrees)	&							& 	 (pc)			\\
\hline
$+$			& $286\degree$	& $331\degree$	&	$5.7 \pm 2.7$			&	$300 \pm 100$	\\
$-$			& $286\degree$	& $331\degree$	&	$7.4 \pm 1.3$			&	$250 \pm 30$	\\
\hline
\\
\end{tabular}\\
Far Carina Arm \\
\begin{tabular}{l c c c r }
\hline
$b$			&	$l_{min}$	&	$l_{max}$	&	$\nicefrac{D}{H_{n_e^2}}$	& 	$H_{n_e^2}$ 	\\
	($\pm$)	&	(degrees)	&	(degrees)	&							& 	 (pc)			\\
\hline
$+$			& $282\degree$	& $292\degree$	&	$6.2 \pm 0.5$			&	$1000 \pm 200$	\\
$-$			& $282\degree$	& $292\degree$	&	$7.3 \pm 0.2$			&	$1000 \pm 200$	\\
$+$			& $292\degree$	& $302\degree$	&	$6.3 \pm 0.7$			&	$1600 \pm 200$	\\
$-$			& $292\degree$	& $302\degree$	&	$5.8 \pm 1.1$			&	$1900 \pm 200$	\\
$+$			& $302\degree$	& $312\degree$	&	$6.8 \pm 1.3$			&	$2100 \pm 400$	\\
$-$			& $302\degree$	& $312\degree$	&	$4.7 \pm 2.0$			&	$2700 \pm 1100$	\\
$+$			& $312\degree$	& $320\degree$	&	$5.41 \pm 0.04$			&	$2600 \pm 300$	\\
$-$			& $312\degree$	& $321\degree$	&	$13 \pm 4$				&	$1700 \pm 700$	\\
\hline
\\
\end{tabular}\\
Scutum-Centaurus Arm \citep{Hill2014}\\
\begin{tabular}{l c c c r }
\hline
$b$			&	$l_{min}$	&	$l_{max}$	&	$\nicefrac{D}{H_{n_e^2}}$	& 	$H_{n_e^2}$ 	\\
($\pm$)		&	(degrees)	&	(degrees)	&							& 	 (pc)			\\
\hline
$+$			& $320\degree$	& $340\degree$	&	$10.7\pm0.4$			&	$330\pm40$		\\
$-$				& $320\degree$	& $340\degree$	&		$13.1\pm0.5$		&	$370\pm30$		\\
\hline
\\
\end{tabular}\\
Perseus Arm \citep{Haffner1999}\\
\begin{tabular}{l c c c r }
\hline
$b$			&	$l_{min}$	&	$l_{max}$	&	$\nicefrac{D}{H_{n_e^2}}$	& 	$H_{n_e^2}$ 	\\
($\pm$)		&	(degrees)	&	(degrees)	&							& 	 (pc)			\\
\hline
$-$			& $125\degree$	& $152\degree$	&		$4.91\pm0.39$		&	$400\pm30$		\\
\hline \hline
\\
\end{tabular}\\
Local \citep{Gaensler2008,Savage2009}\\
\begin{tabular}{l  c }
\hline
	$H_{n_e}$ & 	$H_{n_e^2}$ \\
	 	& (constant $f\left(z\right)$)	\\
	 	 (pc)		&  (pc)	\\
\hline
	$1410^{+260}_{-210}$		&	$700^{+130}_{-100}$	\\
\hline \hline
\\
\end{tabular}\\
\caption{Summary table of results across the Sagittarius-Carina arm, Scutum-Centaurus arm, and Perseus arm. The first column denotes the region above ($+$) or below ($-$) the midplane. Reported values of $H_{n_e^2}$ for the near Sagittarius, near Carina, and far Carina arms are the median ($\pm$ median absolute deviation from the median) across the specified longitude range. The reported $H_{n_e^2}$ for the Perseus arm is modified from the original source to account for more recent distance constraints on the arm segment (see Section \ref{nearsag}) The local $H_{n_e}$ is derived from pulsar Dispersion Measures \citep{Gaensler2008,Savage2009}. If a constant filling fraction, $f\left(z\right)$ is assumed, $H_{n_e} = 2 H_{n_e^2}$. \label{sum}}
\end{center}
\end{table}

\subsection{WIM Filling Fraction} \label{v_dis}

Measures of \halpha~surface brightness provide a measure of the column of free electrons along the line of sight via the emission measure (EM), 

\begin{equation}
EM = \int_{0}^{\infty} n_e^2\left(l\right) dl.
 \label{eq_em}
\end{equation}

\noindent Our determined scale heights are for the distribution of $n_e^2$ ($H_{n_e^2}$), as opposed to $n_e$ ($H_{n_e}$) because of this relationship. $H_{n_e^2}$ and $H_{n_e}$ are related by introducing a volume filling fraction, $f\left(z\right)$ and a characteristic electron density, $n_{c}\left(z\right)$, where 

\begin{equation}
n_e\left(z\right) = f\left(z\right) n_{c}\left(z\right)
 \label{eq_fill}
\end{equation}

and $n_e(z)$ is the space-averaged mean electron density at height $z$. Assuming a constant filling fraction, $f\left(z\right)$, the scale height of the WIM is $H_{n_e} = 2 H_{n_e^2}$ \citep{Haffner1999}. However, the WIM fills more of the volume (relative to neutral components of the ISM) at increasing height, $z$, so a constant filling fraction may not be accurate \citep{Kulkarni1987,deAvillez2000,Hill2012}. For more discussion on adopted filling fraction models and scale heights of the WIM, see \citet{Gaensler2008,Savage2009}. Further study is needed on characterizing the filling fraction as a function of height to properly determine a scale height for the distribution of free elections, $H_{n_e}$.

\section{\stwo~/ \halpha~Line Ratio} \label{ratio_stat}

Preliminary \stwo~emission line data is used to analyze the physical conditions of the WIM around the Carina arm. The \stwo~/ \halpha~line ratio has been shown to increase with decreasing \iha~and increasing distance from the midplane in our Galaxy \citep{Haffner1999} and others \citep{Rand1998, Tullmann2000}. The enhanced forbidden line intensity more closely correlates with \iha~than with distance from the midplane around the Scutum-Centaurus arm, suggesting the observed emission is originating from free electrons in the WIM \citep{Hill2014}. 

To test this relationship around the Carina arm, the \stwo~/ \halpha~line ratio is modeled as a function of \iha~and height, $\left|z\right|$. We assume the line ratio and \iha~follow a power law \citep{Hill2014},

\begin{equation}
\frac{\text{\stwo}}{\text{\halpha}} \propto \left( \text{\iha} \right)^{-a}.
 \label{eq_iha}
\end{equation}

\begin{figure*}[htb!]
\label{ratio_iha}
\epsscale{1.15}
\plotone{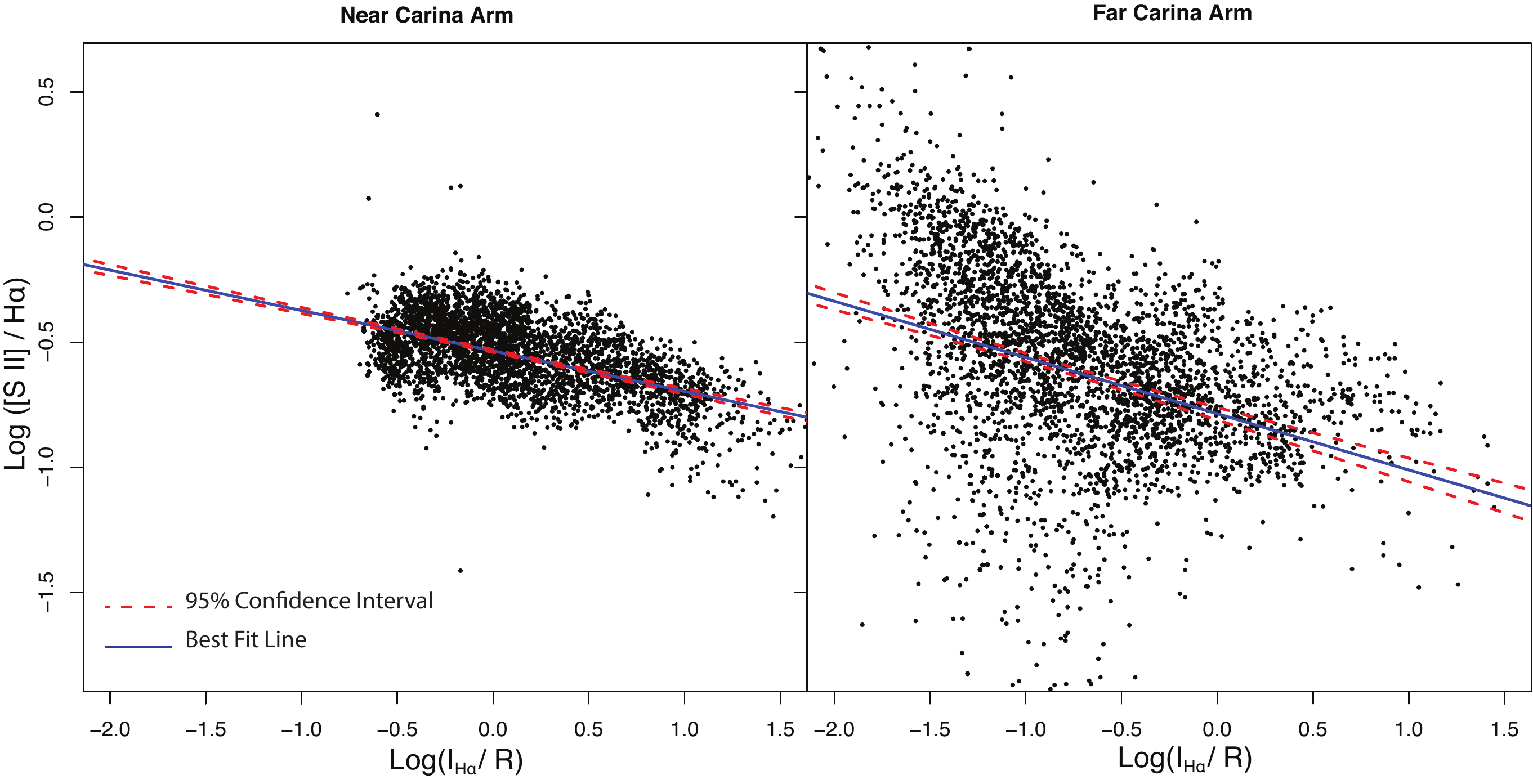}
\caption{\stwo~/ \halpha~line ratio as a function of \iha~along the near (left) and far (right) Carina arm. The solid blue line is the result of a robust least-squares regression fitting of Equation \ref{eq_iha} to the data and the dashed red lines enclose a $95\%$ confidence band for the true regression line. }
\end{figure*}

A least-squares minimization fit is performed to the data (see Figure \ref{ratio_iha}) \citep{r}. Data points correspond to observations ranging from $-0.4 < \text{\tanb}<0.5$ ($-22\degree < b < 27\degree$). The resulting parameters are $a = 0.162 \pm 0.004$ for the near Carina arm and $a = 0.22 \pm 0.01$ for the far Carina arm. 

We assume a linear relationship for the line ratio and $\left|z\right|$ \citep{Hill2014},

\begin{equation}
\frac{\text{\stwo}}{\text{\halpha}}  = \left( \frac{\text{\stwo}}{\text{\halpha}}   \right)_0 + \frac{H_{n_e^2}}{H_{\text{\stwo}}} \text{\tanb}
 \label{eq_tanb}
\end{equation}

\noindent where $\left( \nicefrac{\text{\stwo}}{\text{\halpha}}   \right)_0$ is the midplane value of the line ratio, $H_{n_e^2}$ is the EM scale height, and $H_{\text{\stwo}}$ is the scale height of \stwo~emission. Since the distance to the arm segment, $D$, significantly changes as a function of longitude, this equation is modified to a linear relationship with height, $\left|z\right| = D~\text{\tanb}$, 

\begin{equation}
\frac{\text{\stwo}}{\text{\halpha}}  = \left( \frac{\text{\stwo}}{\text{\halpha}}   \right)_0 + \frac{H_{n_e^2}}{H_{\text{\stwo}}} \frac{1}{D} \left| z \right|.
 \label{eq_z}
\end{equation}

\begin{figure*}[htb!]
\label{z_near}
\epsscale{1.2}
\plotone{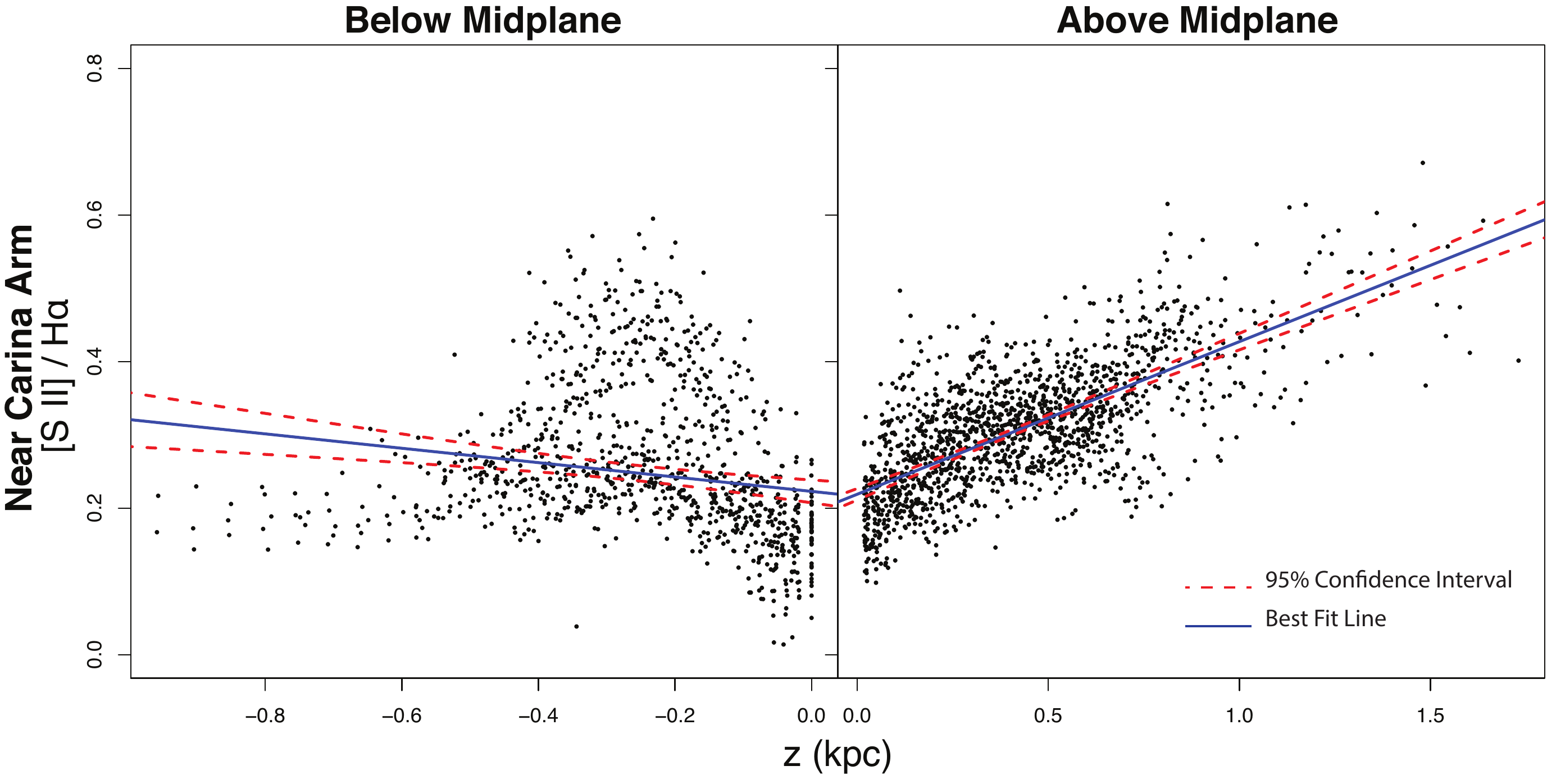}
\caption{\stwo~/ \halpha~line ratio as a function of $z$ for the near portion of the Carina arm. The left and right panels isolate emission below and above the midplane, respectively, assuming a midplane location of $b = 0\degree$. The solid blue line is the result of a robust least-squares regression fitting of Equation \ref{eq_z} to the data and the dashed red lines enclose a $95 \%$ confidence band for the true regression line. Observed behavior below the plane is much less clear, and additional observations of \stwo~further below the plane will help better constrain this relationship. }
\end{figure*}

\begin{figure*}[htb!]
\label{z_far}
\epsscale{1.2}
\plotone{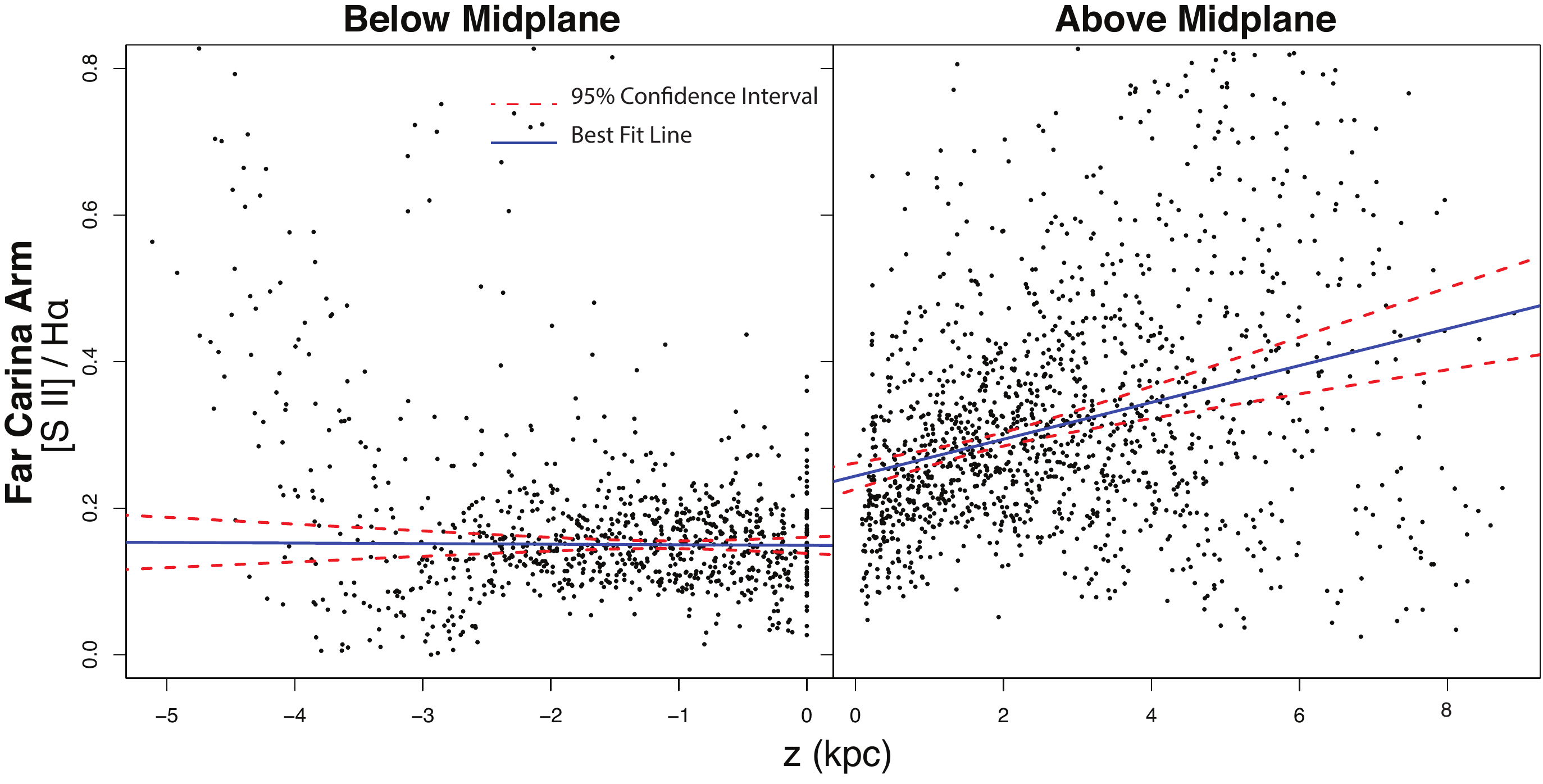}
\caption{\stwo~/ \halpha~line ratio as a function of $z$ for the far portion of the Carina arm. The left and right panels isolate emission below and above the midplane, respectively, assuming a midplane location of $b = 0\degree$. The solid blue line is the result of a robust least-squares regression fitting of Equation \ref{eq_z} to the data and the dashed red lines enclose a $95 \%$ confidence band for the true regression line. Observed behavior below the plane is much less clear, and additional observations of \stwo~further below the plane will help better constrain this relationship. }
\end{figure*}

Four separate least-squares minimization fits are made to the data in Figure \ref{z_near} and Figure \ref{z_far}, separating the data by near and far portions of the arm, and by distance above and below the midplane. Robust regression fitting methods accounting for outliers in the data are used via the robustbase package in R \citep{robust}. The resulting parameters for the near Carina arm are $\left( \nicefrac{\text{\stwo}}{\text{\halpha}} \right)_0 = 0.219 \pm 0.003 \left( 0.223 \pm 0.006 \right)$ and $\nicefrac{H_{n_e^2}}{H_{\text{\stwo}} D} = 0.208 \pm 0.007 \left( 0.10 \pm 0.02 \right)$ above (below) the midplane. The resulting parameters for the far Carina arm are $\left( \nicefrac{\text{\stwo}}{\text{\halpha}} \right)_0 = 0.244 \pm 0.007 \left( 0.149 \pm 0.004 \right)$ and $\nicefrac{H_{n_e^2}}{H_{\text{\stwo}}} = 0.025 \pm 0.004 \left( 0.001 \pm 0.004 \right)$ above (below) the midplane.




Partial correlation tests using the ppcor package in R \citep{ppcor} disentangle the interplay between the line ratio, \iha~and $\left|z\right|$. Many previous studies \citep[see][]{Collins1998, Urquhart2013, Hill2014} use the Spearman method for this test, but Kendall's $\tau$ is a more statistically robust measure of monotonic relationships between variables. Kendall's $\tau$ allows for an exact p-value to be calculated even when ties in the data are present, which is not the case with the Spearman method. Both methods yield very similar results. The partial correlation test analyzes the monotonic correlation between two variables, given a third variable. In this case, the correlation between the line ratio \stwo~/ \halpha~and \iha~($\left|z\right|$) given $\left|z\right|$ (\iha) is analyzed. 

Table \ref{stats} shows the results of these partial correlation tests for the near and far portions of the arm, and for the portions above and below the midplane separately. Our null hypothesis is that of no correlation ($\tau =0$) using a threshold of $p = 0.01$. P-values for the test of correlation between the line ratio \stwo~/ \halpha~and \iha~given $\left| z \right|$ are all well below this threshold, so we reject the null hypothesis of no correlation between the line ratio and \iha~for both the near and far portions of the Carina arm. Above the midplane, p-values for the test of correlation between the line ratio \stwo~/ \halpha~and $\left| z \right|$ given \iha~are also below this threshold, so we reject the null hypothesis of no correlation between the line ratio and $\left| z \right|$ for the near and far Carina arm. However, below the midplane, p-values for the near and far portions are $p = 0.063$ and $p = 0.470$. We fail to reject our null hypothesis for no correlation between the line ratio and $\left| z \right|$ below the midplane of the near and far Carina arm. 

Data below the midplane is more limited, making it harder to see the linear relationship seen above the midplane. However, all values of Kendall's $\tau$ are larger for the correlation between the line ratio and \iha~than the correlation between the line ratio and $\left| z \right|$. Combined, these results show the \stwo~/ \halpha~line ratio is more closely correlated with \iha~than with distance from the plane. The relationship between the line ratio and height is primarily the result of the inverse relationship between \iha~and height.

\begin{table}[htb!]
\begin{center}
Partial Correlations \\
\begin{tabular}{l c c c c r}
\hline
Arm 		& 	b			&	$\tau$ 	& p-value 				& 	$\tau$	&	p-value \\
Segment 	& 	($\pm$)		&	 \iha		& \iha 				& 	$\left|z\right|$		&	$\left|z\right|$		 \\
\hline
Near 		& 		$+$		&$-0.148$ 	& $\expn{1.58}{-17}$		&	$0.048$ 	& $\expn{5.65}{-03}$ \\
			&  		$-$		& $-0.460$ 	& $\expn{1.40}{-102}$		&	$0.040$ 	& \textbf{0.063} \\
Far	 		& 		$+$		&$-0.204$ 	& $\expn{6.13}{-32}$		&	$0.101$	& $\expn{6.74}{-09}$ \\
			&  		$-$		& $-0.169$ 	& $\expn{3.65}{-15}$		&	$0.016$ 	& \textbf{0.470} \\
\hline \hline
\\
\end{tabular}\\
\caption{Summary table of partial correlation tests run with the Kendall $\tau$ statistic. The null hypothesis is that of no correlation (i.e. $\tau = 0$).  The $\pm$ indicates whether the partial correlation test was run on the data above ($+$) or below ($-$) the midplane. A \textbf{bold} p-value indicates that we fail to reject the null hypothesis of no correlation at a threshold of $p = 0.01$. Larger values of $\left| \tau \right|$ indicate a stronger correlation between the variables.  \label{stats}}
\end{center}
\end{table}


\section{Discussion} \label{disc}

\subsection{The Far Carina Arm} \label{disc_far}
Observing faint emission from the far portion of the Carina arm, which lies at large heliocentric distances ($>\unit[10]{kpc}$) is inherently difficult. WHAM observations clearly detect emission at positive LSR velocities in the fourth quadrant of the Milky Way, where most observed emission is expected to be at negative LSR velocities. In the observed directions, this emission most closely corresponds with the CO l-v track of the far Carina arm from \citet{Reid2016}. However, it is difficult to be certain if the observed faint emission is from this spiral structure or from a local expanding bubble or high velocity cloud. The CO track is known not to follow a constant Galactocentric radius and similar \halpha~emission is not seen beyond the Carina arm tangency point, ruling out the possibility of a ring structure. The following sections show a list of features/tests we use to argue for and against the emission corresponding to a distant spiral structure. The pros tend to outweigh the cons, and possible explanations to account for the reasons against are provided. 

\textbf{1. Observed receding perspective effect.}  Although distances to the far Carina arm are not well constrained, the heliocentric distance to the spiral arm should increase significantly as a function of Galactic longitude beyond the tangency point. The \halpha$~$and \stwo$~$maps of the far Carina arm clearly show this perspective effect (see solid white lines of constant height, $z$ in Figure \ref{maps}).   

\begin{figure*}[htb!]
\label{axisym}
\epsscale{1}
\plotone{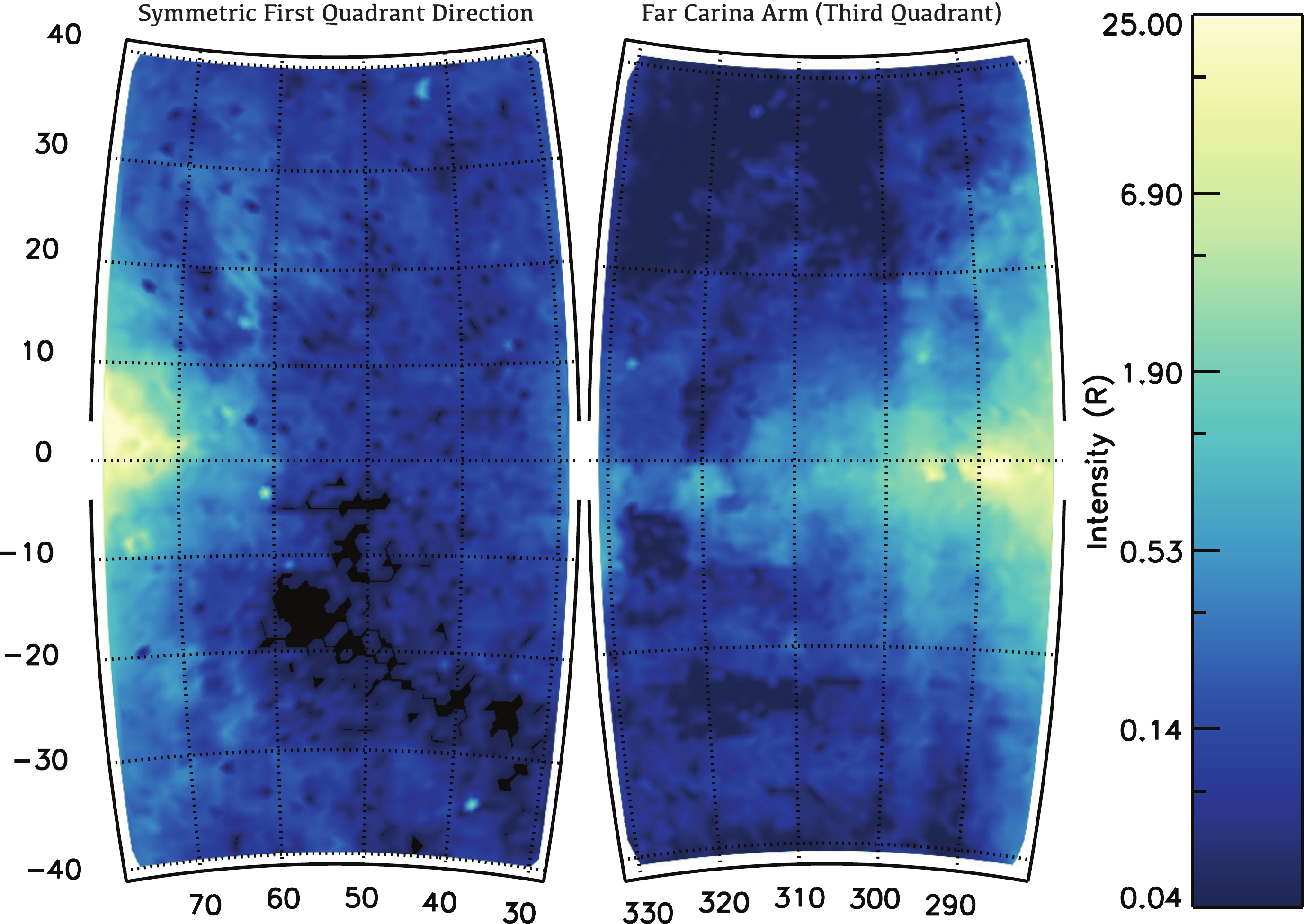}
\caption{Map of \iha~showing the non-axisymmetric nature of the observed far Carina arm structure. The right side shows the far Carina arm in the third quadrant of the Galaxy, as seen in Figure \ref{maps}, and the left side shows the equivalent map in the first quadrant of the Galaxy. Data are integrated following the same l-v track and velocity width used for the far Carina arm, but with longitude reflected to the first quadrant and velocity reflected to negative values. These maps are significantly different.}
\end{figure*}

\textbf{2. Non-axisymmetric structure.} It is well-understood that spiral structures in galaxies are not symmetric across the minor axis. To check for this, we make a symmetric map along the first quadrant, where negative LSR velocities kinematically imply $R_G > R_{\odot}$. WHAM data is integrated along the same l-v track used in the fourth quadrant, but instead longitude values are reflected to the first quadrant and the sign of the velocity is reflected to negative values. The resulting map, along with the original map in the fourth quadrant are shown in Figure \ref{axisym}. The two maps are distinct, and the first quadrant does not show a similar structure in \halpha~emission.

\textbf{3. Symmetric across the midplane.} Spiral arms are aware of the midplane of the Galaxy. The \halpha$~$and \stwo$~$emission is generally symmetric in Galactic latitude as seen in the maps of Figure \ref{maps} and Figure \ref{mapss} and in the Galactic latitude profiles of \halpha$~$emission plotted in Figure \ref{fits}. 

\textbf{4. Location of midplane in Galactic latitude.} Figure \ref{fits} shows the midplane of the Galaxy at the near and far Carina arm are at distinctly different latitudes, with the near arm leaning towards a midplane location of $b > 0\degree$ and the far arm towards $b < 0\degree$. A single Gaussian fit to the intensity profile as a function of Galactic latitude locates the midplane. This increases our confidence in separating the near and far components of the arm along a single line of sight. The distinct behavior of these two physical regions help rule out the possibility of observing an extended wing of closer, negative velocity gas reaching out to positive velocities in the fourth quadrant. The LAB survey \citep{Kalberla2005} shows the midplane location of the neutral gas in the far Carina arm also tends towards $b < 0\degree$ following the same method. Additionally, at the large distances associated with the far Carina arm, the dust lane in the plane would appear to be very narrow (within one WHAM beam). The intensity as a function of height still shows exponential behavior down to low latitudes close to $b \approx 0\degree$.

\textbf{5. Peaks in emission spectra. } Figure \ref{bv_peak} shows a sample of peaks of spectra decomposed into Gaussian components, shown in blue. Local maxima of the observed spectra are shown in red. Seeing local maxima correspond to peaks in the Gaussian components increases our confidence that the observed emission feature is from a real Galactic source, rather than a wide wing from local emission at $v_{LSR} \approx \unit[0]{km~s^{-1}}$.

\textbf{6. Reverse distance argument.} Our method provides a direct measure of the ratio of distance to the arm segment, $D$, to the EM scale height, $H_{n_e^2}$. If we assume the scale height of the far Carina arm is consistent with other spiral arm segments (see Table \ref{sum}), then we can estimate a distance to the arm segment based on our measured values. Using a "nominal" value of $H_{n_{e^2}} = \unit[0.3]{kpc}$, the average distance is $D \approx \unit[2]{kpc}$.  This places the far Carina arm at around the same distance as the near Carina arm. However, this "reverse distance argument" is only valid if the scale height of the WIM around all spiral arm segments is constant and consistent with previous measurements. The Sagittarius-Carina arm is much different than both the Perseus and Scutum-Centaurus arms, so a direct comparison may not be valid. The Carina arm shows up much more clearly in gas diagnostics \citep{Cohen1985, Dame2007, Reid2016} than through counts of old stars, in which the Perseus and Scutum-Centaurus arms stand out \citep{Benjamin2009}. This difference suggests the Perseus and Scutum-Centaurus arms have different potential wells than the Carina arm and contain more dense gas and more star formation activity. An explanation of the correlations between our anomalous scale height measurements and other oddities in the stellar and molecular gas structure of the Carina arm requires further study and is beyond the scope of this paper.

\textbf{7. CO and \halpha$~$velocity offset.} Our l-v track for the Carina arm is defined using CO emission but the velocity centroids of the \halpha$~$spectra for the far Carina arm are offset from the CO emission to slightly more positive velocities (see Figure \ref{bv_peak}). However, there is no reason to believe the cold molecular components of the ISM are spatially and kinematically coincident with the \halpha~emitting WIM. Other galaxies, such as M$51$ show such a physical offset between CO gas and diffuse ionized gas \citep{Schinnerer2013}. This offset could be consistent with where star formation takes place within a spiral arm. Additionally, the trends in the \halpha~emission as a function of longitude are still closely correlated with the CO data.

There is clearly-detected emission at velocities that most closely lie near the expected velocities for the far Carina arm, and a negative velocity counterpart is not seen as would be expected for an expanding local bubble. Based on these arguments and tests, we see strong evidence for the emission originates in a distant spiral arm. 

\subsection{Physical Conditions}

The \stwo~/ \halpha~line ratio is known to show large variations while still having a strong correlation with \halpha~intensity \citep{Haffner2009}. The narrow confidence intervals despite the large scatter in Figure \ref{ratio_iha} illustrates this. A power-law between the \stwo~/ \halpha~line ratio and \iha~is well supported along the Carina arm (see Figure \ref{ratio_iha}), along with local gas, the Perseus arm, and the Scutum-Centaurus arm \citep{Haffner1999, Hill2014}. The most likely physical explanation for this relationship is a change in the temperature of the gas. However, this attribution is not straightforward, unlike with the \ntwo~/ \halpha~line ratio which strongly correlates with the temperature of the gas \citep{Haffner1999, Madsen2006, Haffner2009}. 

\citet{Otte2002} showed the line ratio depends on many physical conditions:

\begin{equation}
\frac{\text{\stwo}}{\text{\halpha}}  = \left( 7.49 \times 10^5 \right) T_4^{0.4}~e^{\nicefrac{2.14}{T_4}}~\frac{\text{H}}{\text{H}^+}~\frac{\text{S}}{\text{H}}~\frac{\text{S}^+}{\text{S}}
 \label{eq_ratio}
\end{equation}

\noindent where $T_4$ is the temperature of the emitting gas (in units of $\unit[10^4]{K}$), $\nicefrac{\text{H}^+}{\text{H}}$ and $\nicefrac{\text{S}^+}{\text{S}}$ are the hydrogen and sulfur ionization fractions, and $\nicefrac{\text{S}}{\text{H}}$ is the sulfur abundance. Following previous work \citep{Hill2014}, we assume the sulfur abundance does not undergo large variations in the ISM, and adopt the \citet{Reynolds1998} result for a hydrogen ionization fraction of $\nicefrac{\text{H}^+}{\text{H}} \gtrsim 0.9$ in the WIM. Then, variations in the line ratio are either from changes in the gas temperature or changes in the sulfur ionization fraction. 

In-depth photoionization modeling of sulfur would disentangle this relationship, but these models are difficult to construct due to a poorly constrained temperature dependence of the dielectric recombination rate of sulfur \citep{Ali1991,Barnes2014}. Instead, previous studies \citep{Haffner1999, Madsen2006, Haffner2009} show variations in the \stwo~/ \halpha~line ratio largely track variations in temperature (larger \stwo~/ \halpha~$\implies$ larger $T_4$) by showing strong correlation with variations in the \ntwo~/ \halpha~line ratio. Meanwhile, the \stwo~/ \ntwo~line ratio traces changes in the sulfur ionization state. Future work will incorporate ongoing \ntwo~observations with WHAM to fully understand the temperature distribution of the WIM throughout the arm.  

Following this reasoning and the results of statistical tests (see Section \ref{ratio_stat}), we conclude the temperature of the emitting gas (as traced by the \stwo~/ \halpha~line ratio) is more strongly correlated with in-situ electron density (as traced by \iha~and the EM) than with height above/below the midplane ($\left|z\right|$). These results further support the conclusions of \citet{Hill2014}: the heating mechanisms involved for the WIM at different $n_e$ do not vary greatly with $\left|z\right|$. Mechanisms other than photoelectric heating are necessary \citep[such as cosmic ray heating, magnetic reconnection, dissipation of turbulence, photoelectric emission from dust grains;][]{Reynolds1992,Reynolds1999,Otte2002, Barnes2014, Rand1998, Wiener2013}.

This behavior further suggests \iha~more closely correlates with electron density in the emitting gas, rather than distance from the plane. The observed enhancement in the line ratio at larger heights is strong evidence for the observed \halpha~emission originating from the diffuse in-situ gas of the WIM, as opposed to primarily consisting of scattered light from higher density \ion{H}{2} regions as modeled by \citet{Seon2012}. Their model suggests an increase in distance from O and B stars in \ion{H}{2} regions explains the observed line ratio enhancement. However, the Sagittarius-Carina arm and the Scutum-Centaurus arm \citep{Hill2014} suggest the line ratio more closely depends on the gas density as opposed to height. In our preferred picture, the \stwo~/ \halpha~line ratio is tracing variations in temperature and gas density within the WIM and not distance from OB stars as expected by scattered light models. 

\section{Conclusions and Summary} \label{summary}

WHAM is used to study the structure and physical conditions of the warm $\left( \approx \unit[8000]{K} \right)$ ionized medium throughout the Sagittarius-Carina arm. Faint emission from the spiral arm is kinematically isolated using a CO-informed longitude-velocity track. Emission is detected across a large range in Galactic longitude $(\text{near Sagittarius; }20\degree < l < 52\degree \text{ near/far Carina; } 282\degree < l < 332 \degree )$, spanning a large range in Galactocentric radii $(\unit[6]{kpc} \lesssim R \lesssim \unit[11]{kpc})$ and heliocentric distance $(\unit[1]{kpc} \lesssim D \lesssim \unit[15]{kpc})$. The \halpha~intensity as a function of height above and below the plane suggest an EM scale height ($H_{n_{e}^2} \approx \unit[300]{pc}$) consistent with other spiral arms in the Galaxy (Scutum-Centaurus, Perseus) for the near Sagittarius and near Carina arms. Emission seen around the far Carina arm suggest significantly larger scale heights ($H_{n_{e}^2} > \unit[1000]{pc}$). 

The anomalously large scale heights along the far Carina arm suggest a significant physical difference in the environment surrounding this region of the Galaxy. We offer a few potential explanations for the large scale heights observed, but a complete explanation requires a more in-depth study of the trends in other ISM components  within and around this distant spiral arm. At large Galactocentric radii, the scale height of the WIM in the far Carina arm tends to increase rapidly with increasing radius. However, as seen in Figure \ref{summary}, the Perseus arm has a much smaller scale height at similar Galactocentric radii. 
Independent measures of the scale height of ionized gas from pulsar dispersion measures \citep{Gaensler2008, Savage2009} seem to match well with the observed trend for the increasing scale height as a function of Galactocentric radius along the Sagittarius-Carina arm (see Figure \ref{summary}).

The large scale heights for the far Carina arm suggest the star formation rate is much different here than within the Perseus or Scutum-Centaurus arm, but there is not direct evidence or measurements of the star formation activity along the far reaches of this arm. Further study of the differences of the midplane conditions and star formation along the far Carina arm are beyond the scope of this work. Future work will incorporate ongoing \hbeta~emission observations from WHAM to better analyze the ionized gas conditions near the midplane.

Statistical analysis of the \stwo~/ \halpha~line ratio throughout the near and far Carina arm show a stronger correlation between the line ratio and \halpha~intensity (inverse power law), as opposed to height, $z$, above the midplane (linear). This supports the interpretation for the \stwo~/ \halpha~line ratio primarily tracing variations in gas temperature and density. We interpret this as evidence for the observed diffuse \halpha~emission ($\gtrsim 80\%$) originating from in-situ gas in the WIM, as opposed to scattered light from \ion{H}{2} regions. 

Future work will use \ntwo~emission line observations from WHAM to more closely analyze the temperature behavior of the WIM around this spiral arm structure. Ongoing observations of \hbeta~will allow for accurate dust extinction corrections throughout all observed lines of sight. 

\acknowledgments
We acknowledge the support of the U.S. National Science Foundation (NSF) for WHAM development, operations, and science activities. The survey observations and work presented here were funded by NSF awards AST-0607512 and AST-1108911. R.A.B. acknowledges support from the NASA grants NNX10AI70G and HST-GO-13721.001-A.  A.S.H acknowledges support from NASA grant HST-AR-14297.001-A. We would also like to acknowledge the support of NSF REU Site grant AST-1004881 and the contributions of undergraduate researchers Peter Doze (Texas Southern University), Andrew Eagon (University of Wisconsin-Whitewater), and Alex Orchard (University of Wisconsin-Madison) to the initial analysis of the WHAM all-sky survey results.

\bibliographystyle{aasjournal}
\bibliography{apj-jour,Carina_Arm}






%

\end{document}